\pdfoutput=1
\documentclass[prb,twocolumn,showpacs,preprintnumbers,amsmath,amssymb,floatfix]{revtex4}
\usepackage{graphicx}
\usepackage{bm}
\usepackage{float}

\newcommand \be{\begin{eqnarray}}
\newcommand \ee{\end{eqnarray}}
\newcommand \ba{\begin{align}}
\newcommand \eea{\end{align}}
\newcommand {\ket}[1]{|#1\rangle}
\newcommand {\bra}[1]{\langle #1|}

\newcommand {\p}[1]{\partial_{#1}}

\newcommand \V{\vec}
\newcommand \uu{\uparrow}
\newcommand \dd{\downarrow}

\begin{document}
           \csname @twocolumnfalse\endcsname
\title{Kinetic theory of spin-polarized systems in electric and magnetic fields with spin-orbit coupling: II. RPA response functions and collective modes}
\author{K. Morawetz$^{1,2,3}$ 
}
\affiliation{$^1$M\"unster University of Applied Sciences,
Stegerwaldstrasse 39, 48565 Steinfurt, Germany}
\affiliation{$^2$International Institute of Physics (IIP)
Federal University of Rio Grande do Norte
Av. Odilon Gomes de Lima 1722, 59078-400 Natal, Brazil
}
\affiliation{$^{3}$ Max-Planck-Institute for the Physics of Complex Systems, 01187 Dresden, Germany
}

\begin{abstract}
The spin and density response functions in the random phase approximation
(RPA) are derived by linearizing the kinetic equation including a magnetic
field, the spin-orbit coupling, and mean fields with respect to an external
electric field. Different polarization functions appear describing various
precession motions showing Rabi satellites due to an effective Zeeman
field. The latter turns out to consist of the mean-field magnetization, the
magnetic field, and the spin-orbit vector. The collective modes for charged
and neutral systems are derived and a threefold splitting of the spin waves
dependent on the polarization and spin-orbit coupling is shown. The dielectric
function including spin-orbit coupling, polarization and magnetic fields is
presented analytically for long wave lengths and in the static limit. The dynamical screening length as well as the long-wavelength dielectric function shows an instability in charge modes, which are interpreted as spin segregation and domain formation. The spin response describes a crossover from damped oscillatory behavior to exponentially damped behavior dependent on the polarization and collision frequency. The magnetic field causes ellipsoidal trajectories of the spin response to an external electric field and the spin-orbit coupling causes a rotation of the spin axes. The spin-dephasing times are extracted and discussed in dependence on the polarization, magnetic field, spin-orbit coupling and single-particle relaxation times. 
\end{abstract}
\maketitle
%

\section{Introduction}

The development of spintronic devices is largely based on the understanding of collective spin waves.
Spin waves, besides density waves, are one of the fundamental collective
excitations in strongly interacting Fermi systems, e.g., in ferromagnetic
materials \cite{PGN05,VWJ14}, graphene \cite{SAHR11,TN13}, or isospin
excitations in nuclear matter \cite{BPP94}. In the past, this had motivated
people to develop Green functions techniques for the quantum transport theory of spin resonance \cite{LW72,SLW76}. 

If the range of interaction is shorter than the DeBroglie wavelength, such
excitations are also predicted \cite{LeRi68,Ba81,LhLa82a,GuMu84,Ba86,RuLe89}
and observed \cite{JoDe84,BiFr89} in dilute spin-polarized gases. The
transverse spin-wave dynamics has been the subject of a series of theoretical investigations \cite{Ba86,MR05}. 
In ultracold gases, spin waves have been observed, even spatially resolved
\cite{LHWC02,GLHNWC02}, and were described by longitudinal spin waves
\cite{WNCC02}. The spin diffusion in trapped Bose gases shows an anisotropy in
modes \cite{MR06} and the collisionless damping has been seen to deviate for
quadrupole modes from  experiments \cite{NWC02} indicating the role of
collisional correlations. The spin-wave damping has been measured in a
polarized Fermi-liquid-like $^3$He-$^4$He mixture even at zero temperature
\cite{PV06} and as an 'identical spin-rotation effect' \cite{NTCL84}.

The influence of the magnetic field on such spin waves is of special interest,
e.g., as magneto-transport effects in paramagnetic gases \cite{Bas91}. The
Landau levels influence the spin relaxation \cite{BL04,E06}, which has been
measured with the help of spin coherence times \cite{SLMGFA04}. The influence
of magnetic fields is treated in various systems ranging from plasma
\cite{CKB79}, solid-state plasmas \cite{HY86}, and semiconductors \cite{LMT01}
to spin-orbit coupled systems \cite{B05} and graphene \cite{Shiz07}. The
feedback of magnetization dynamics due to spins on the spin dynamics itself is
reviewed in Ref. \cite{LSLMKL13}. The Zeeman field is reported to trigger a
transition from a charge density wave to a spin density wave \cite{AGRVL08}. Quite promising for technological applications turns out the possibility to create magnetic nanooscillations by pure spin currents \cite{DUG14}. The spin current can be converted into a terahertz electromagnetic pulse due to the inverse spin Hall effect \cite{KBMENMZFMBWROM13}.

Quite recently the spin-orbit coupling has moved to the center of interest
\cite{W03,WJW10} since this coupling allows to convert spin waves into spin
currents, which is important for spintronic devices \cite{CHINTKBF15}. There
has been observed a spin-orbit-driven ferromagnetic resonance
\cite{FKWVZCCGJF15} which shows that an effective magnetic field is created in
the magnetic material by oscillating electric currents. This is also the basis
of microwave spectroscopy \cite{LLU13}. Earlier this has been identified as a
magneto-electric effect where a charge current induces a spin polarization
known as the Edelstein effect \cite{E90,BGVR14}. Spin-polarized longitudinal currents can be induced due to spin-orbit interaction in certain crystal symmetries \cite{WSCKE15}. Experimentally even a planar Hall effect has been reported using spin waves \cite{KMSSH15} as well as spin polarization oscillations without spin precession \cite{BKF14}.

Coulomb interactions are known to reduce the effect of spin-orbit coupling in the spin-Hall effect \cite{Hu:2003}. The phonon-modulated spin-orbit interaction has been investigated to show that the screening is influenced by the spin-orbit coupling \cite{GF97}. Screening effects play a crucial role for the temperature dependence of conductivity in quasi two-dimensional systems \cite{Mm02} , monolayer graphene \cite{A06} and multilayer graphene \cite{G07}. The Coulomb correction to the conductivity in graphene had covered an involved debate \cite{GCCNS13,JVH10,HJVO08,M08}. With this respect the extraction of correct spin relaxation or dephasing times has been in the center of interest \cite{LFHGIRFSH07,LHHSFR07,LHHSFR08} since it is most promising for new storage devices.

During the last few years, many researchers have shown an appreciable interest
in the dielectric function and the properties of screening in two-dimensional
gases with spin-orbit coupling \cite{PG06,BAVF09,SSS12}. Similar results
appear if the pseudospin response function in doped graphene is calculated
\cite{WC07,BPAM09,PPV09,G11}. The random phase approximation (RPA) is
calculated in this respect for single and multilayer graphene
\cite{YRK11,PAPM12}. These approaches calculate the Lindhard dielectric
function with form factors arising from chirality subbands. Additional energy
denominators appear if four-band approximations are considered \cite{G11}
where a band gap appears \cite{TR12}. The comparison of pristine graphene,
Dirac cones, and gaped graphene with an antidot lattice can be found in Ref. \cite{SJP11}. These responses are needed if one wants to understand the optical properties of graphene irradiated by an external electric field \cite{BPJ12}. 

All of these approaches consider the spin degree of freedom as an inner
property of particles leading to form factors in the Lindhard dielectric
function. Here, the spin degree of freedom is considered on equal statistical
footing with the particle distribution leading to more forms of the response
function due to the spin-orbit coupling, satellites, and Zeeman splitting by
magnetic fields and self-energy effects that cannot be cast into a Lindhard
form with form factors. This has an impact on the collective density and spin
modes. The will calculate analytically the threefold splitting of spin modes
\cite{UF10} as a function of the spin-orbit coupling and the effective Zeeman field.  

In this paper, we want to present a unifying treatment of density and spin
waves in the random phase approximation including the spin-orbit coupling,
magnetic fields, and an arbitrary magnetization for systems with charged and
neutral scattering. This will allow us to investigate the influence of
spin-orbit coupling on the screening properties of the Coulomb interaction as well as the collective modes in systems with neutral scattering. For this purpose, we linearize the kinetic equation derived in the first part of this paper. Linearizing the mean-field kinetic equation yields the RPA response since a lower-level kinetic equation provides a response of higher-order many-body correlations \cite{Mc02}.

Following a short summary of the basic kinetic equation derived in the first
part of the paper, the linear response to an external electric field is
presented in the second section. This results into coupled equations for the
spin and density response with a variety of dynamical polarization functions
describing different precessions. In Sec. III, the charge and spin density
response functions are analyzed with respect to their collective modes and the
spin waves are discussed for neutral and charged scattering. The polarization
causes a splitting of spin modes. For certain polarizations and spin-orbit
coupling, an instability occurs, which is interpreted as spin-domain
separation. This is underlined by the influence of spin-orbit coupling and
polarization on the screening properties in charged systems where the
instability occurs in spatial domain. The dielectric function including the
spin-orbit coupling and an effective medium-dependent Zeeman field is derived
analytically in the long-wavelength and static limits respectively. The
dynamic and static screening lengths are discussed there. The spin response
due to an applied electric field is then extracted and the spin-dephasing
times are discussed. As in the Edelstein effect, the applied electric field
causes a charge current and a spin response that shows oscillations dependent
on the spin-orbit coupling, magnetic field, and relaxation time. In Sec. V, we
present the linear response including arbitrary magnetic fields and show how
the normal Hall and quantum Hall effects appear from the kinetic theory. As an
important point, a subtlety in retardations due to the magnetic field is
presented. A summary finishes this second part of the paper. In Appendix, some
useful expressions for solving involved vector equations are presented.

Let us now shortly summarize the
quantum kinetic equation derived in the first part of the paper.
We describe the density and polarization density by their corresponding Wigner functions
\be 
\sum_p f(\V x,\V p,t)=n(\V x,t),\qquad
\sum_p \V g(\V x,\V p,t)=\V s(\V x,t)
\label{quasin}
\ee
where $\sum\limits_p=\int d^Dp/(2\pi\hbar)^D$ for $D$ dimensions.
As a result of the first part of this paper, the four Wigner functions
\be
\hat \rho=f+\V \sigma\cdot \V g=
\begin{pmatrix}
f+g_z &g_x-i g_y\cr g_x+i g_y&f-g_z
\end{pmatrix}
\label{rhofg}
\ee
have been shown to obey coupled kinetic equations
\be 
D_t f+\V A \cdot \V g &=&0\nonumber\\
D_t \V g+\V A f
&=&2 (\V \Sigma\times \V g)
\label{kinet}
\ee
where $D_t=(\p t+\V {\cal F}\V {\p p}+\V v\V {\p x})$ describes the drift and force of the scalar and vector part with the velocity 
\be
v={p\over m_e}+\p p \Sigma_0
\label{vel}
\ee
and the effective Lorentz force
\be
\V {\cal F}=(e \V E +e \V v \times  \V B - \V {\p x}
\Sigma_0).
\label{lor}
\ee
This effective Lorentz force as well as the velocity both become modified due to the scalar meanfield selfenergy
\be
\Sigma_0(\V q,t)&=&
n(\V q,t) V_0(\V q)+\V s(\V q,t)\cdot \V V(\V q)
\ee
as a spatial convolution between
the density and spin polarization with the Fourier transformed scalar and
vector potentials, respectively. The latter ones originate from magnetic
impurities and/or effective magnetizations in the material. Here, we
concentrate on the intrinsic spin-orbit coupling. The meanfields with extrinsic spin-orbit coupling are given in III.C of the first part of the paper.

The second parts on the left side of (\ref{kinet}) represent the coupling between the spin parts of the Wigner distribution given by the vector drift
\be
A_i=(\V \partial_p \Sigma_i\V \partial_x-\V \partial_x\Sigma_i\V \partial_p+e(\V \partial_p\Sigma_i \times \V B)\V \partial_p).
\ee
We subsumed in the vector selfenergy
\be
\V \Sigma=\V \Sigma_{H}(\V x,t)+\V b(\V x,\V p,t)+\mu_B \V B(\V x,t)
\label{sig}
\ee
the magnetic impurity meanfield
\be
\V \Sigma_H=n(\V q,t) \V V(q)+\V s(\V q,t) V_0(q),
\ee 
and the spin-orbit coupling vector $\V b$, as well as the Zeeman term $\mu_B
\V B$ such that the effective Hamiltonian possesses the Pauli structure
\be
H_{\rm eff}=H+\V \sigma\cdot \V \Sigma
\ee
with the effective scalar Hamiltonian
\be
H={k^2\over 2 m_e}+\Sigma_0(\V x,\V k,T)+e \Phi(\V x, T)
\ee
where ${\V k}={\V p}-e {\V A}({\V x},t)$ ensures gauge invariance. 
Any spin-orbit coupling found in the literature can be recast into the form $\V \sigma \cdot \V b(p)$ as illustrated in Table I of the first part of this paper.
The vector part of (\ref{kinet}) finally contains additionally the spin-rotation term on the right-hand side responsible for spin precession.

\section{Linear Response}

\subsection{Without magnetic field but conserving relaxation time} 

Let us consider the linearization of the kinetic equation (\ref{kinet}) with
respect to an external electric field, no magnetic field and in a homogeneous situation. We Fourier transform the time $\partial_t\to -i\omega$ and the spatial coordinates $\V \partial_x\to i \V q$. 
The Wigner functions are linearized according to $\hat \rho(\V x,\V p,t)=f(\V p)+\delta f(\V x,\V p,t)+\V \sigma \cdot
[\V g(\V p)+\delta \V g(\V x,\V p,t)]$ due to the external electric field perturbation
$e \delta \V E=e \V E(\V x,t)=-\nabla \Phi$. The density and spin-density variation reads 
\be
\delta n(\V q,\omega)&=&\sum\limits_p\delta f(\V q,\V p,\omega)\nonumber\\
\delta \V s(\V q,\omega)&=&\sum\limits_p \delta \V g(\V q,\V p,\omega)
\ee
and the density and spin-density linear response functions are given by
\be
\delta n(\V q,\omega)&=&\chi(\V q,\omega) \Phi\nonumber\\
\delta \V s(\V q,\omega)&=&\V \chi^s(\V q,\omega) \Phi.
\label{res}
\ee

Further, we assume a collision integral of the relaxation time approximation \cite{HaMo92}
\be
-\frac 1 2 [\hat \tau^{-1}, \delta \hat \rho^l]_+
\ee
with the vector and scalar parts of the relaxation times
$\hat \tau^{-1}=\tau^{-1}+\V \sigma \cdot \V \tau^{-1}$ where
\be
\tau={\tau^{-1}\over \tau^{-2}-|\V \tau^{-1}|^2},\quad \V \tau=-{\V \tau^{-1}\over \tau^{-2}-|\V \tau^{-1}|^2}.
\ee

In the first part of the paper, the relaxation of the kinetic equations (\ref{kinet}) has been shown towards the two-band distribution $f={f_++f_-\over 2}$ and $\V g=\V e \,\,{  f_+-f_-\over 2}$ with the Fermi-Dirac distribution 
\be
f_0(\epsilon_p(r)\pm|\V \Sigma(p,r)|)
\label{twob}
\ee 
and the precession direction $\V e=\V \Sigma/\Sigma$. Now we assume a relaxation towards a local distribution $f^l=f_0(\epsilon \pm |\Sigma|-\mu-\delta \mu)$ such that the density conservation can be enforced \cite{Mer70,D75},
\be
\delta n&=&\sum\limits_p (f-f_0)=\sum\limits_p (f-f^l+f^l-f_0)
\nonumber\\
&=&\sum\limits_p (f^l-f_0)=\partial_\mu n \,\delta \mu,
\ee
as expressed by the second line.
Therefore the scalar relaxation term becomes
\be
-{\delta \hat \rho^l\over \tau}=-{\delta \hat \rho\over \tau}+{\delta n\over \tau\partial_\mu n}\partial_\mu \hat \rho_0.
\ee
In this way the density is conserved in the response function which could be
extended to include more conservation laws \cite{Ms01,Ms12}. If we consider
only the density conservation but not the polarization conservation, we can
restrict ourselves to the $\p \mu f_0$ term. Please note that we neglect in
this way the interference effects of disorder \cite{ZNA01}.

Abbreviating now $-i \omega+i \V p\cdot \V q/m+ \tau^{-1}=a$ and $iq\p p\V \Sigma+\V \tau^{-1}=\V {\cal B}$, 
the coupled kinetic equations (\ref{kinet}) take then the form
\be
a \delta f+\V {\cal B} \delta \V g&=&S_0\nonumber\\
a \delta \V g+\V {\cal B} \delta f-2 \V \Sigma\times \delta \V g&=&\V S
\label{kineq}
\ee
with $e\V E=-i \V q \Phi$ and
\ba
S_0&=iq\p p f (\Phi+\delta \Sigma_0)+i q \p p \V g \cdot \delta \V \Sigma+{\delta n\over \tau\partial_\mu n}\partial_\mu f_0\nonumber\\
{\V S}&=iq\p p \V g (\Phi+\delta \Sigma_0)+i q\p p f \delta \V \Sigma+2 (\delta \V \Sigma \times \V g)+{\delta n \partial_\mu \V g\over \tau\partial_\mu n}.
\label{source}
\end{align}
In order to facilitate the vector notation we want to understand $q\p p=\V q \cdot \V{\p p}$ in the following. 

\subsection{Collective modes from balance equations}

Multiplying the linearized kinetic equations (\ref{kineq}) with powers of momentum and integrating one obtains coupled hierarchies of moments. 
A large variety of treatments neglect certain Landau-liquid parameters
\cite{Mi04} based on the work of Ref. \cite{Le70} in order to close such system. A
more advanced closing procedure was provided by Ref. \cite{Mi05} where the energy dependence of $\delta \V s$ was assumed to be factorized from space and direction $\V p$ dependencies.

We will not follow these approximations here but solve the linearized equation exactly to provide the solution of the balance equations and the dispersion. Amazingly, this yields a quite involved and extensive structure with much more terms than usually presented in the literature. Nevertheless, it is instructive to have a first look at the balance equation for the densities
\ba
\p t \delta n+\p {x_i} \V J_i+\V \tau^{-1}\cdot \delta \V s&=0
\nonumber\\
\p t \delta \V s+\p {x_i} \V S_i+\V \tau^{-1} \delta n-2\sum\limits_p\V \Sigma \times \delta \V g &=2\delta \V \Sigma \times \V s
\label{balancens}
\end{align}
where we Fourier transformed the wavevector $q$ back to spatial coordinates $x$. Then the density currents and magnetization currents 
\ba
\hat J_j=\frac 1 2 \sum\limits_p [\hat \rho, v_j]_+
&=\sum\limits_p\left [
f v_{j}\!+\!\V g \cdot \partial_{p_j} \V b \!+\!\V \sigma\cdot (v_{j}\V g\!+\!f \partial_{p_j} \V b )
\right  ]\nonumber\\
&=J_j+\V \sigma\cdot \V S_j.
\label{current}
\end{align}
appear exactly as expected from the elementary definitions, see Sec. III.G of
part I.

We are now interested in the long wavelength limit $q \to 0$, which means we neglect any spatial derivative in (\ref{balancens}). Alternatively, we might consider this as the spatially integrated values providing the change of number of particles and magnetization
\be
N=\int d^3x n(x)=n_{q=0},\quad \V m=\int d^3x \V s(x)=\V s_{q=0}.  
\label{mN}
\ee
The first equation of (\ref{balancens}) gives in frequency space
\be
-i\omega \delta N+\V \tau^{-1}\cdot \delta \V m=0
\ee
with the help of which we get the closed equation for the magnetization from (\ref{balancens})
\ba
&-i\omega \delta \V m\!-\!{i\over \omega} (\V \tau^{-1}\!+\!2 \V m\times \V V)\V \tau^{-1}\cdot \delta \V m\!-\!2 (N\V V\!+\!\mu_B \V B)\times \delta \V m
\nonumber\\
&=2 \sum\limits_p\V b(p)\times \delta \V g_{q=0}
.
\label{balancem}
\end{align}
In the right-hand side, 
all terms that are coming from the explicit knowledge of the solution $\delta
\V g$ needed to evaluate this sum over the momentum-dependent spin-orbit term
$\V b$ are collected. The separation of the balance equation in this form has the merit to see already the collective spin mode structure. The fine details are then worked out when we know the explicit solution in Sec. \ref{coll}.

Since we have $\V m=m\V e_Z$ and $\V V=V \V e_Z$ the equation for the magnetization becomes 
\ba
\!\begin{pmatrix}\!\!
-i\omega&2 (\!N V\!\!+\!\!\mu_B B)&0\cr
-2 (\!N V\!\!+\!\!\mu_B B)&-i\omega & 0\cr
0& 0& -i\omega\!
\end{pmatrix} \!\delta \V m\!=\!2 \!\sum\limits_p\V b\!\times \! \delta \V g_{q=0}
\label{mode0}
\end{align}
neglecting the quadratic terms of the vector relaxation times $\V \tau^{-1}$. The latter would add a term
$-\!i (\V \tau^{-1}\!+\!2 \V m\times V)\V \tau^{-1}\cdot \delta \V m/\omega$ to the left-hand side.

Inverting (\ref{mode0}) provides the solution of the magnetization change provided we know the solution of $\delta \V g$ on the right hand side.
Interestingly, this inversion is only possible for a nonzero determinant. The vanishing determinant provides therefore the eigenmodes of spin waves with some possible modifications due to the spin-orbit coupling term. 

The dispersion relation from the condition of vanishing determinant in (\ref{mode0}) yields the two spin waves
\be
\omega_{\rm spin}=\pm 2 |N V+\mu_B B|
\label{coll0}
\ee
which shows the linear splitting due to the driving external magnetic field and the permanent magnetization $\V V=V\V e_z$. 

\subsection{Solution of linearized coupled kinetic equations}

As we have seen, even the balance equations for the linearized kinetic equations (\ref{kineq}) are not closed if we do not know the solution for $\delta \V g$, which comes from the momentum dependence of $\V \Sigma$.
In the following we will present two ways to solve (\ref{kineq}). First the elementary direct way and secondly with the help of operator algebra. The latter is then applicable directly to solve the kinetic equation with magnetic fields in the next section.

\subsubsection{Solution with the help of vector equation }

Solving the first equation of (\ref{kineq}) for 
\be
\delta f=-{1\over a}\left (\V {\cal B} \cdot \delta \V g-S_0 \right )
\label{soldeltaF0}
\ee
and introducing the result into the second equation leads to a vector equation of type 
(\ref{3e}) 
and with the help of the abbreviations
\be
\V c=-{\V {\cal B}\over a},\quad \V z=\frac{\V \Sigma}{a},\quad
\V o=\frac 1 a
\left (\V c S_0+\V S\right )
\ee
and setting $B=\V o$, $A=2\V z$, and $Q=-V=\V c$ it can be readily solved 
[ see Eq. (\ref{3s})], which becomes
\ba
&\delta \V g=
{\V o\!+\!\V c\!\times\! (\V c \!\times\! \V o)\!+\!2 (\V z\!\times\! \V o)\!+\!4 \V z (\V z\cdot \V o)\!-\!2 (\V z\cdot \V c)(\V c \!\times\! \V o)\over 1-c^2+4 z^2-4(\V z\cdot \V c)^2}
\nonumber\\
&\delta f=\V c \cdot \V \delta g+{S_0\over a}.
\label{soldeltaF}
\end{align}

\subsubsection{Solution with the help of operator algebra}

As a second possibility, we rewrite equations (\ref{kineq}) with the help of the identity 
\ba
&\V c\cdot \V g+ (\V \sigma \cdot \V c) f-2 \V \sigma \cdot (\V \sigma\times \V g)=
\nonumber\\&
\V \sigma \cdot {\V c+2 i\V \sigma\over 2} \hat \rho+\hat \rho \V \sigma \cdot {\V c-2 i\V
  \sigma\over 2}
\label{identity}
\end{align}
into one operator equation for $\delta \hat F=\delta f+\V \sigma \cdot \V \delta g$ and transform the frequency back in time $-i \omega\to \p t$:
\ba
&\hat S_p(t)=S_0(t)\!+\!\V\tau\cdot \V S(t)=\left (\p t+i{pq\over m}+\tau^{-1}\right )\delta \hat F(t)
\nonumber\\&\!+\!\left({\V b\over 2}\!+\!i \V \Sigma\right)\cdot \V \sigma \, \delta \hat F\!+\!\delta \hat F \left({\V b\over 2}\!-\!i \V \Sigma\right)\cdot \V \sigma.
\label{opS}
\end{align}
This equation is easily solved
\ba
&\delta \hat F(t)=\!\!\int \limits_{-\infty}^t \!\!\!d\bar t\, {\rm e}^{i({pq\over m}\!-\!{i\over \tau})(\bar t \!-\!t)}
{\rm e}^{\left({\V b\over 2}\!+\!i \V \Sigma\right)\cdot \V \sigma(\bar t\!-\!t)} 
\hat S_p(\bar t) 
{\rm e}^{\left({\V b\over 2}\!-\!i \V \Sigma\right)\cdot \V \sigma(\bar t\!-\! t)}
\nonumber\\&
\end{align}
and transformed back in frequency space to obtain
\ba
\delta \hat F(\omega)=\int\limits_{-\infty}^0 \!d x\, {\rm e}^{i\left ({pq\over m}-\omega-{i\over \tau}\right )x}
{\rm e}^{\left({\V b\over 2}\!+\!i \V \Sigma\right)\cdot \V \sigma x} 
\hat S_p(\omega) 
{\rm e}^{\left({\V b\over 2}\!-\!i \V \Sigma\right)\cdot \V \sigma x}.
\nonumber\\&
\label{251}
\end{align}
Further evaluation is presented in Appendix~\ref{operator}. With the help of
(\ref{c11}) and (\ref{c12}) we evaluate the corresponding integrals and all scalar terms determine $\delta f$, and all terms proportional to $\sigma$ determine $\V \delta g$. We obtain again the result (\ref{soldeltaF}) \footnote{There is a puzzling difference in the denominator. We have instead of the square of scalar product $(\V c \cdot \V z)^2$ now $c^2 z^2$ which difference is just the squared cross product. Since we restrict later to linear orders in $q\p p$, i.e. linear orders in $\V c$ we can neglect this difference. 
}
\ba
&\delta \V g=
{\V o\!+\!\V c\!\times\! (\V c \!\times\! \V o)\!+\!2 (\V z\!\times\! \V o)\!+\!4 \V z\cdot (\V z\cdot \V o)\!-\!2 (\V z\cdot \V c)(\V c \!\times\! \V o)\over (1-c^2)(1+4 z^2)}.
\label{soldeltaF1}
\end{align}
To see the known limits, we inspect some further approximations.

\subsubsection{Long-wavelength limit}
We assume only a scalar relaxation time and neglect therefore the vector part
describing skew scattering and side jump effects. Further we use the long
wavelength limit and expand in first orders of $q\p p \V\Sigma$ which
translates into $\V c^ 2\approx 0$ in (\ref{soldeltaF1}). Here, we have abbreviated the form of the mean-field self-energy
\be
U&=&\Phi+\delta \Sigma_0
\nonumber\\
\V U&=&\delta \V \Sigma
\ee
and we used $e \V E=-i \V q \Phi$, $\bar \omega=i a=\omega-\V p\cdot \V q/m+ i \tau^{-1}$ and the meanfields 
\be
\delta \Sigma_0&=&
V_0 \delta n+\V V \cdot \delta \V s,\quad
\delta \V \Sigma =
\V V \delta n+V_0 \delta \V s
.
\label{deltaS}
\ee
Then the solution (\ref{soldeltaF0}) and (\ref{soldeltaF1}) reads $[\V g=g \V e$]
\ba
\delta f&=-\frac{1}{\bar \omega} \biggl \{U q\p p f+\V U q \p p \V g-i{\delta
  n\over \tau \p \mu n} \p \mu f-q\partial_p \V \Sigma\cdot \V \delta g \biggr \}
\nonumber\\
\delta \V g&=\delta \V g^{\rm asy}+\delta \V g^{\rm sym}_{\rm Rabi}+\delta \V g^{\rm sym}_{\rm n}
\label{solkin}
\end{align}
Each of the terms corresponds to one of the three possible precession motions: 
\be
{1\over 4 |\Sigma|^2-\omega^2}\begin{pmatrix} -i \omega \cr 
{4 i|\Sigma|^2\over \omega} \cr 2 |\Sigma| \end{pmatrix}
=\int\limits_0^\infty {\rm e}^{i\omega t} \begin{pmatrix} \cos 2|\Sigma| t\cr 1-\cos 2|\Sigma| t \cr \sin 2|\Sigma t|\end{pmatrix}.
\label{costime}
\ee

The $\sin 2 |\Sigma| t$ motion is responsible for the anomalous Hall effect
and their terms are collected in $\delta \V g^{\rm asy}$. Let us write out the explicit forms. The symmetric solution consists of an frequency denominator with Rabi-satellites
\be
&\delta \V g^{\rm sym}_{\rm Rabi}=\frac{1}{2} \left ({1\over \bar \omega -2\Sigma}+{1\over \bar \omega -2\Sigma}\right )
\left [
-U g q \p p \V e-2 i \V g\times \delta \V \Sigma
\right .\nonumber\\
&\left .+q\p p f\V e \times (\V e\times \V U)+i{\delta n\over \tau \p \mu n}g\p \mu \V e \right ]
\label{ds1}
\ee
and a part with a normal denominator $\bar \omega=\omega+i/\tau-\V p\V q/m$:
\ba
\delta \V g^{\rm sym}_{\rm n}={1\over \bar \omega}
\left [
-U \V e q \p p g -q\p p f\V e \times (\V e\cdot \V U)+i{\delta n\over \tau \p \mu n}\p \mu g\V e \right ]
\label{ds2}
\end{align}
which can be combined together
\ba
&\delta \V g^{\rm sym}=\frac{1}{\bar \omega(4 |\V \Sigma|^2-\bar \omega^2)} 
\biggl \{
[\bar \omega^2 \V U-4 |\V \Sigma|^2 \V e (\V e \cdot \V U)] q\p p f
\nonumber\\
&\qquad +
\bar \omega^2 (U q \p p \V g-i{\delta n\over \tau \p \mu n}\p \mu \V g)
\nonumber\\
&\qquad 
-4 \Sigma^2 (U q \p p g-i{\delta n\over \tau \p \mu n}\p \mu g)\V e 
-2 i \bar \omega^2 \V U \times \V g
\biggr \}.
\end{align}

The term responsible for the anomalous Hall effect reads
\ba
&\delta \V g^{\rm asy}=
\frac{i}{2}  \left ({1\over \bar \omega +2\Sigma}-{1\over \bar \omega -2\Sigma}\right )
\biggl \{\V e \times \V U q \p p f\!
\nonumber\\
&+\left [U \V e \!\times\! q\p p \V e \!-\!2 i \V e \!\times\! (\V U \!\times\! \V e)-i{\delta n\over \tau \p \mu n}\V e\times \p \mu \V e \right ] g
\biggr \}
\label{ds3}
\end{align}

In order to compare with the homogeneous solution presented in Eq. (143) of part I
\be
\delta \V \rho(\omega,k)&=& 
{i \omega \over 4 |\Sigma|^2-\omega^2} e E\partial_k\V g
\nonumber\\&&
-4 i{1\over \omega (4 |\Sigma|^2-\omega^2)}\V \Sigma (\V \Sigma \cdot eE\partial_k \V g)
\nonumber\\&&
-2  {1  \over 4 |\Sigma|^2-\omega^2} \V \Sigma \times eE\partial_k \V g.
\label{deltarho}
\ee 
we take the $q\to 0$ limit of (\ref{solkin}) with $\V q U=ie\V E+o(q), q \V U=o(q)$ and obtain
\ba
\delta f=&-{i\over \bar \omega} \left (e E \p p f-{\delta n\over \tau \p \mu n} \p \mu f\right )
\nonumber\\
\delta \V g^{\rm sym}=&-{i\bar \omega \over \bar \omega^2-4 |\Sigma|^2} 
\left (e E \p p \V g-{\delta n\over \tau \p \mu n} \p \mu \V g -2\V U \times \V g\right )
\nonumber\\
&+{4 i\Sigma^2\over \bar \omega (\bar \omega^2-4 |\Sigma|^2)}
\left (e E \p p g-{\delta n\over \tau \p \mu n} \p \mu g\right )\V e
\nonumber\\ 
\delta \V g^{\rm asy}=& {2 |\Sigma| g\over \bar \omega^2-4 |\Sigma|^2}
\left [\V e \times e E \p p \V e -2 \V e\times (\V U\times \V e)
\right .
\nonumber\\ 
&\left . \qquad \qquad \qquad-{\delta n\over \tau \p \mu n}{\V e}\times \p \mu
  \V e \right ].
\label{qsol}
\end{align}
Without vector meanfield variation $\V U\approx 0$ and relaxation time, the
last term responsible for the anomalous Hall effect corresponds directly to
the third one in Eq. (\ref{deltarho}). The first two terms correspond to the
first two ones in Eq. (\ref{deltarho}) as simple algebra shows observing that
$\V \Sigma \cdot \partial \V e=0$ since $\V e =\V \Sigma/|\V \Sigma|$, $\V e
(\V e\cdot \partial \V g)=\V e \partial g$ and $\V e \times (\V
e\times \partial \V g)=-g\partial \V e$. The term $\V e\times eE\partial_p \V
e$ of Eq. (\ref{qsol}) corresponds to the precession term found in
Ref. \cite{B05} as an additional rotation of the magnetization.

\subsection{Response functions}

We want now to integrate the linearized solution (\ref{ds1}), (\ref{ds2}) and
(\ref{ds3}) over the momentum to obtain the density and spin response
functions including the meanfield and spin-orbit coupling effects. This will
lead to a selfconsistent equation. We can design the dimensionality of the
considered problem as done above after Eq. (\ref{quasin}). 

The final result for the particle and spin density response using only intrinsic meanfields $\delta \Sigma_0=V_0 \delta n+\V V\cdot \delta \V s$ and $\delta \V \Sigma=\V V \delta n+V_0 \delta \V s$ leads to the following linear system  
\ba
&(1-\Pi_0 V_0-\V \Pi \cdot \V V+{i\Pi_{0\mu}\over \tau\p \mu n})\delta n=\Pi_0 \Phi
+(\Pi_0\V V+\V \Pi V_0) \cdot \delta \V s
\nonumber\\&
(1-\Pi_0 V_0-\overleftrightarrow \Pi V_0)\delta \V s=\V \Pi_3 \Phi 
+\V \Pi_3(\V V \cdot \delta \V s)
+V_0\V \Pi_2\times \delta \V s
\nonumber\\
&
+(V_0 \V \Pi_3+\Pi_0 \V V+\V \Pi_2 \times \V V+\overleftrightarrow \Pi \cdot \V V-{i\over \tau\p \mu n}\V \Pi_\partial)\delta n
\label{final}
\end{align}
with the abbreviations for the polarizations
\be
\V \Pi_2&=&\V \Pi_g+\V \Pi_{xf}
\nonumber\\
\V \Pi_3&=&\V \Pi+\V \Pi_{xg}+\V \Pi_{e}
\nonumber\\
\V \Pi_\partial&=&\V \Pi_{x\mu}+\V \Pi_{\mu}+\V \Pi_{\mu e}
\nonumber\\
\overleftrightarrow \Pi&=&\overleftrightarrow \Pi_{fe}+\overleftrightarrow \Pi_{xe}.
\label{abbrev}
\ee
The $1/\tau$ terms come from the Mermin conserving relaxation-time approximation which means a relaxation towards a local equilibrium specified such that local conservation laws are obbeyed \cite{Mer70,D75,Ms12}. 

It was helpful here to define the polarization functions according to the three precessions expressed by the parts (\ref{ds1}), (\ref{ds2}) and (\ref{ds3}).
The standard polarization functions for scalar and vector distribution coming from (\ref{ds1}) read with $\bar \omega=\omega-\V p\cdot \V q/m+i/\tau$
\be
\Pi_{0}(q\omega)&=&-\sum\limits_p q \p p f {1\over \bar \omega}
\nonumber\\
\Pi_{0\mu}(q\omega)&=&-\sum\limits_p \p \mu f {1\over \bar \omega}
\nonumber\\
\V \Pi(q\omega)&=&-\sum\limits_p q \p p \V g {1\over \bar \omega}
\nonumber\\
\V \Pi_\mu(q\omega)&=&-\sum\limits_p \p \mu \V g {1\over \bar \omega}.
\ee

The remaining parts of (\ref{ds1}) and (\ref{ds2}) combine into the forms of  
\be
{4\Sigma^2\over \bar \omega (4 \Sigma^2-{\bar \omega}^2)}={1\over \bar \omega}-\frac 1 2 \left ({1\over \bar \omega+2\Sigma}+{1\over \bar \omega-2\Sigma}\right )
\ee
which vanish quadratically with the vector meanfield 
\be
\V \Pi_{e}(q\omega)&=&\sum\limits_p gq\p p \V e {4\Sigma^2\over \bar \omega (4 \Sigma^2-{\bar \omega}^2)} 
\nonumber\\
\V \Pi_{\mu e}(q\omega)&=&\sum\limits_p g \p \mu \V e 
{4\Sigma^2\over \bar \omega (4 \Sigma^2-{\bar \omega}^2)}
\nonumber\\
\overleftrightarrow \Pi_{fe}(q\omega)&=&\sum\limits_p q\p p f (1-\V e\circ \V e){4\Sigma^2\over \bar \omega (4 \Sigma^2-{\bar \omega}^2)}
\ee
and a rotation part
\be
\V \Pi_{g}(q\omega)&=&-i  \sum\limits_p \V g\left ({1\over \bar \omega+2\Sigma}+{1\over \bar \omega-2\Sigma}\right ).
\label{Pg}
\ee

The responses from the asymmetric part (\ref{ds3}) lead to 
\be
\V \Pi_{xf}(q\omega)&=&i\sum\limits_p\V e q \p p f\frac 1 2 
\left ({1\over \bar \omega+2\Sigma}-{1\over \bar \omega-2\Sigma}\right )
\nonumber\\
\V \Pi_{xg}(q\omega)&=&i \sum\limits_p\V e \times q \p p \V g\frac 1 2 
\left ({1\over \bar \omega+2\Sigma}-{1\over \bar \omega-2\Sigma}\right )
\nonumber\\
\V \Pi_{x\mu}(q\omega)&=&i \sum\limits_p\V e \times \p \mu \V g\frac 1 2 
\left ({1\over \bar \omega+2\Sigma}-{1\over \bar \omega-2\Sigma}\right )
\nonumber\\
\overleftrightarrow \Pi_{xe}(q\omega)&=&2\sum\limits_p g (1-\V e\circ \V e)
\frac 1 2 
\left ({1\over \bar \omega+2\Sigma}-{1\over \bar \omega-2\Sigma}\right )
\nonumber\\&&
\label{pol1}
\ee
which vanish linearly in orders of the self-energy. In the case of vanishing spin-orbit coupling, i.e. no momentum dependence of $\V e$, we have $\V \Pi_{xg}=\V \Pi_{x\mu}=\V \Pi_{e}=\V \Pi_{\mu e}=0$. 

It is known
that vertex corrections lead to additional structure factors in the RPA 
polarization functions, which would extend the expressions here. Here, it is
shown that the spin-orbit coupling causes a zoo of additional RPA
polarization forms even
on the level of the single-loop approximation, which is the highest level to be
obtained by the linearization of the mean-field equations.

\section{Spin and density waves\label{coll}}

The system of density and spin responses (\ref{final}) allows us to determine the spin and density waves that might be excited in the system. It is convenient to continue to work in the wave number space $r\leftrightarrow q$. For the magnetization and total particle number (\ref{mN}),
we consider only dipole modes that are characterized by the first-order moments
\be
\Psi_j&=&\int d^3r x_j \delta n(r)=-i \p {q_j} \delta n_q |_{q=0}\nonumber\\
\V \Psi_j&=&\int d^3r x_j \delta \V s(r)=-i \p {q_j} \delta \V s_q |_{q=0}.
\label{prop1}
\ee
The linear response for dipole modes is assumed to be characterized by linear deviations $\delta f=\V \beta_1\cdot \V {\p r} f_0+\V \beta_2 \cdot \V {\p p} f_0$
and $\delta g=\V \alpha_1\cdot \V {\p r} g_0+\V \alpha_2 \cdot \V {\p p} g_0$
such that we have
\be
\delta n|_{q=0}&=&i\V \beta_1\cdot \V q n|_{q=0}=0\nonumber\\
\delta \V s|_{q=0}&=&i\V \alpha_1\cdot \V q \V s|_{q=0}=0
\label{prop2}
\ee
for the long-wavelength limit of the deviations themselves.
In order to determine the eigenmodes of (\ref{final}), we do not need the actual values of $\alpha$ and $\beta$. Due to the properties (\ref{prop1}) and (\ref{prop2}), we can apply the $q$-derivative directly to (\ref{final}) and obtain an equation system where $\delta n$ and $\delta \V s$  are replaced by $\Psi_j$ and $\V \Psi_j$ in (\ref{final}) and all response functions have to be taken in the $q\to 0$ limit. Any derivative of the latter vanishes since they are connected with terms (\ref{prop2}). This simplifies the analysis appreciably and shows the strength of the response system (\ref{final}). For quadrupole modes where the second derivative is needed, one has to calculate also the $q\to 0$ limit of the derivatives of the polarization functions.

Further analysis relays on the expansion of different polarization functions as presented in Appendix~\ref{expansion}.
Then, the momentum integration still cannot be performed analytically if the
momentum-dependent spin-orbit term $b(p)$ in $\V \Sigma$ is present. Therefore
we treat them linearly in $\V b(p)$, which allows to give an explicit form. In
fact, these terms are only present if the spin-orbit coupling creates an
explicit $p$-dependence in $\V \Sigma=\V \Sigma_n +\V b(p)$, where we denote
the momentum-independent selfenergy with $\V \Sigma_n=n\V V+V_0 \V s+\mu_B \V
B$. Let us define the $z$-direction by this last term $\V e_z=\V \Sigma_n/|\V \Sigma_n|$ and expand all directions in first order of $\V b(p)$.

The direction of effective polarization becomes
\be
\V e &=& {\V \Sigma\over |\Sigma|}=\V e_z \left ( 1-{b_\perp^2\over 2}\right )+{\V b_\perp}\left (1-{b_3}\right )
\ee
where we use the short-hand notation
\be
{\V b_p\over \Sigma_n}=\V b_\perp+\V e_z b_3.
\label{bsenk}
\ee
Since the distribution functions in equilibrium are functions of $|\V \Sigma|$ according to (\ref{twob}), i.e. functions of $b_\perp^2$ and $b_3$, and since the latter ones are even in momentum direction, the distributions are even in momentum direction as well. Therefore the polarization becomes
\be
\V s=\sum\limits_p g\V e =\V e_z \sum\limits_p g\left (1-{b_\perp^2\over
    2}\right )=\V e_z \left (s_0-{B_g^2\over 2}\right )
\label{s}
\ee
with
\be
s_0&=&\sum\limits_p g;\qquad B_g^2=\sum\limits_p b_\perp^2 g;\qquad m=s_{q=0}.
\label{Bg}
\ee

The $q\to 0$ limit is of course dependent on the $q$-dependence of the
potential. Therefore let us analyze the neutral scattering $V_0$ and charged
scattering $V_0=e^2/\epsilon q^2$ separately. As we will see, only the latter
provides density waves as a collective plasma oscillation.

\subsection{Excitation with neutral scattering}  

To study the excitation modes for scattering with neutral impurities, we can
restrict ourselves to the lowest-order expansion in $q$, which simplifies the results of
the last section again. Especially $\Pi_0=\V \Pi=\V \Pi_{xf}=\V \Pi_{xg}=\V \Pi_e=\overleftrightarrow \Pi_{fe}=0$.
According to (\ref{prop1}) and (\ref{prop2}) we obtain for the density excitation from (\ref{final}) 
\be
\left (1-{i\over \omega_+ \tau}+o(q^2)\right )\Psi_j=0
\ee
with $\omega^+=\omega+i/\tau$ which means that we have either the zero mode $\omega=0$ or $\Psi_j=0$, i.e., no density dipole wave excitations. This will be different when we consider charged scattering in the next section.

With the help of this result the equation for the spin excitation becomes from (\ref{final})
in long-wavelength expansion
simply
\be
{\cal D} \V \Psi_j\equiv (1-V_0 \overleftrightarrow \Pi)\V \Psi_j-V_0\V \Pi_2\times \V \Psi_j=0
\label{300a}
\ee
with
\ba
\V \Pi_2&=i\V e_z
\left [-s_0+{B_g^2\over 2}(1-\Sigma_n\partial_{\Sigma_n})-
{q^2E_g\over D m_e}\partial_\omega^2\right ]{2\omega\over \omega^2-4\Sigma_n^2}
\nonumber\\
\overleftrightarrow \Pi&=
\!\begin{pmatrix} \!s_0\!+\!B_{g11}\!+\!{B_g^2\Sigma_n\over 2}
\partial_{\Sigma_n}&0&0\cr 0&s_0\!+\!B_{g22}\!+\!{B_g^2\Sigma_n\over 2}
\partial_{\Sigma_n}&0\cr 0&0&-B_{g}^2\!\end{pmatrix}
\nonumber\\&\times{4\Sigma_n\over 4 \Sigma_n^2\!-\!\omega_+^2}.
\end{align}
The vanishing determinant of ${\cal D}$ in (\ref{300a}) yields the collective spin excitation wave. 

Neglecting first the thermally averaged spin-orbit terms completely 
we obtain the modes
\be
\omega_{\rm spin}=-{i\over \tau}\pm 
2 |s V_0-\Sigma_n|
\label{modes0}
\ee
which shows that two spin modes are excited. Provided we have an interaction
$V_0$ the mode is shifted simply by the meanfield selfenergy $\Sigma_n=n V+V_0
s+\mu_B B$ and this mode $\omega\sim 2 \mu_B B^{\rm eff}$ is exclusively
dependent on the effective Zeeman shift $\mu_B B^{\rm eff}=n V+\mu_B B$. This
result we had already obtained as zero order in spin-orbit coupling
(\ref{coll0}) in agreement with the recent report of a transition from charge
to spin density waves only appearing at a finite Zeeman field \cite{AGRVL08}.

As discussed in Sec. III.H of part I of this paper [], the selfconsistency would result into the replacement $\Sigma_n=n V+V_0 s+\mu_B B\to (n V+\mu_B B)/(1+{m_e\over 2 \pi \hbar^2} V_0)$. In order to facilitate the following notation we write shortly $\Sigma$ for this selfconsistent $\Sigma_n$.

\begin{figure}[]
\includegraphics[width=8cm]{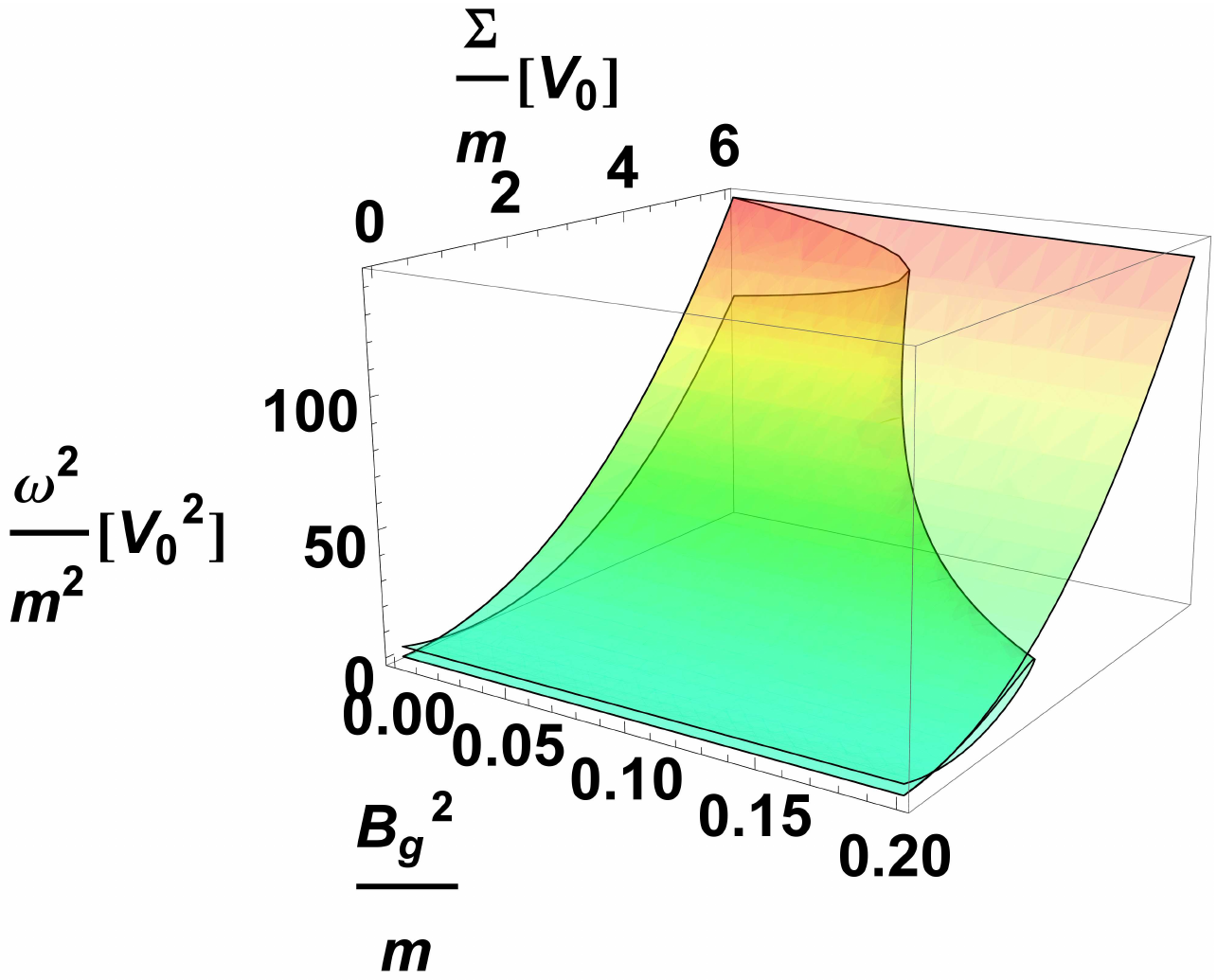}
\caption{\label{mode_neutral} The three frequencies of spin excitation mode (\ref{threemodes}) as a function of selfenergy $\Sigma$ and thermally averaged spin-orbit coupling (\ref{Bg}) and magnetization $m=s_{q=0}$.}
\end{figure}

The dispersion including the spin-orbit coupling is dependent on the 
parameter $B_g^2=B_{g11}^2+B_{g22}^2$ given in (\ref{Bg}).
The spin modes as zeros of $|{\cal D}|$ of (\ref{300a}) appear to be
\ba
&\left (\omega+{i\over \tau}\right )^2-4 (\Sigma-s V_0)^2=2 V_0\left [ B_g^2 \Sigma
\right .\nonumber\\&\left .
+(s V_0-2 \Sigma)\left (s\pm \sqrt{s^2-{2\over V_0} \Sigma B_g^2
}
\right )
\right ]
\label{threemodes}
\end{align}
together with a third mode as the sum of the right side.
The result 
is plotted in Fig. \ref{mode_neutral}. One sees that the two modes
(\ref{modes0}) appear, which differ with increasing $\Sigma$. The threefold
splitting of spin modes was reported in Ref. \cite{UF10}.

A closer inspection shows that one mode can become imaginary for small selfenergies $\Sigma$. In fact, expanding up to $+o(B_\perp^2,)$ one gets
\ba
&\left (\omega_{\rm spin}+{i\over \tau}\right)^2=
4\left \{
\begin{matrix}
\Sigma^2\left (1
-{B_g^2\over s}\right )
\cr
(\Sigma-s V_0)^2+{B_g^2\Sigma(\Sigma-s V_0)\over s}
\end{matrix}
\right .
\label{mode1}
\end{align}
where the square of the first mode can become negative which means an imaginary mode.
Due to the six-order polynomial in $\omega$, for each dispersion, also the complex conjugated one is a solution. A finite imaginary part means instability when it overcomes the damping by collisions $1/\tau$.

\begin{figure}[]
\noindent\parbox[]{8.7cm}{
\parbox[]{4.2cm}{
\includegraphics[width=4.2cm]{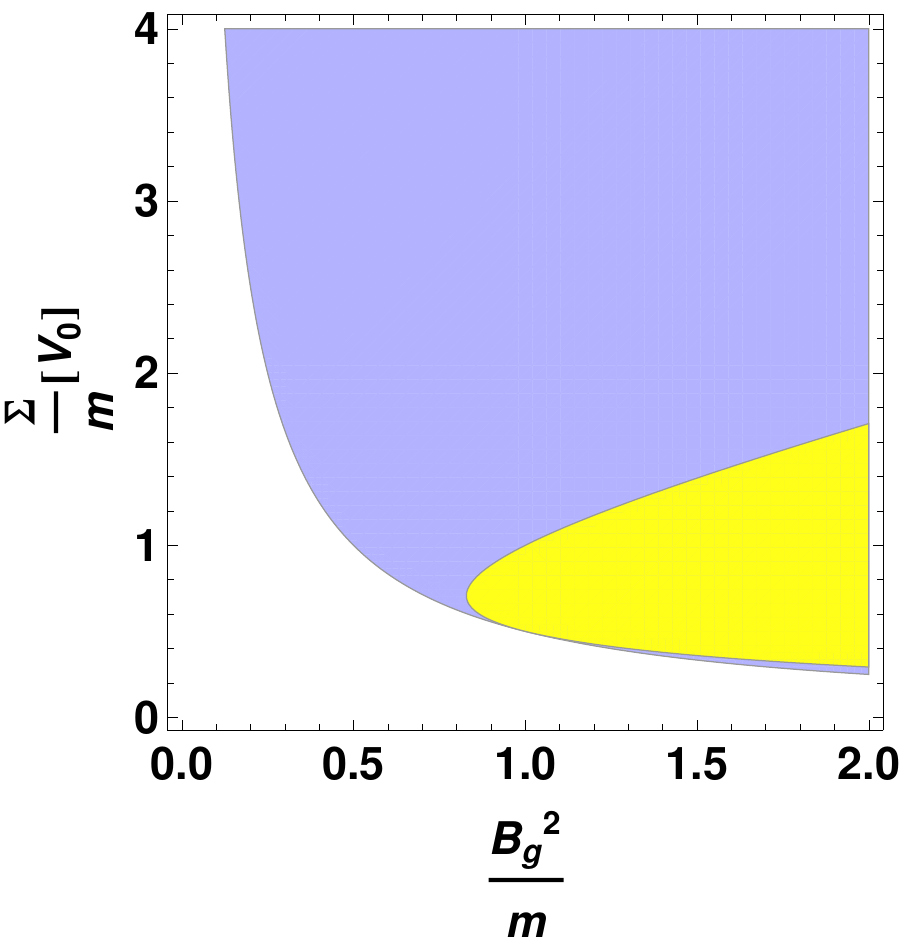}
}\hspace{0.1cm}
\parbox[]{4.3cm}{
\includegraphics[width=4.3cm]{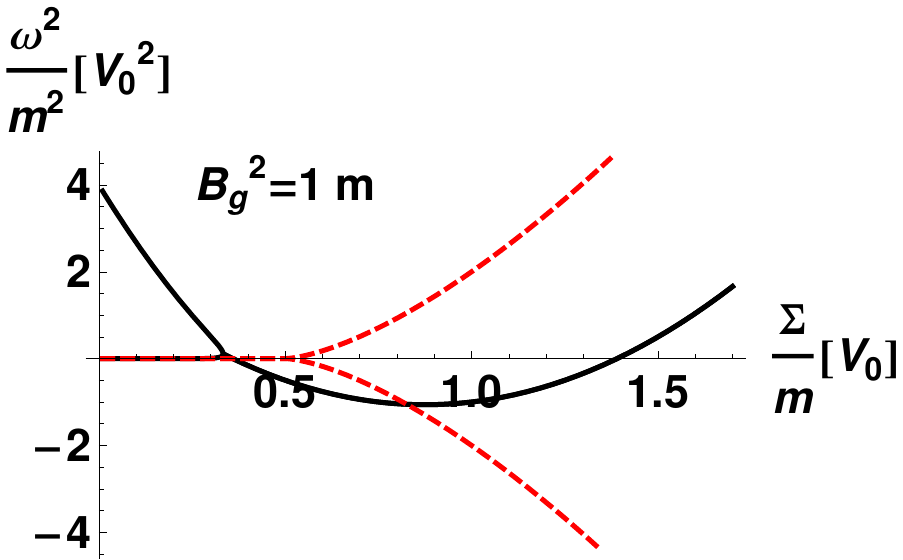}
}}
\caption{\label{mode_neutral_insta} Left: The maximal range of spin-wave instability for vanishing collisional damping. The outer area (blue) designs the ${\rm Im} \omega^2\ne 0$ modes and the inner (yellow) area the $\omega^2<0$ modes. Right: The real (black solid) and imaginary (dashed red) part of $\omega^2$ versus $\Sigma$}
\end{figure}

The maximal range of such possible spin-wave instability (without collisional
damping $1/\tau\to 0$) is shown in Fig. \ref{mode_neutral_insta}. Here, we
distinguish the region of $\omega^2<0$ appearing as the inner region (yellow) and the region where ${\rm Im}\, \omega^2\ne 0$ as the outer (blue) one. The twofold regions of complex frequencies are seen in the cut in Fig. \ref{mode_neutral_insta}.

The physics of these instabilities can be seen more explicitly in the zero
temperature limit where in quasi-two-dimensions and in the presence of linear Dresselhaus $\beta=\beta_D$ or Rashba $\beta=\beta_R$ spin-orbit coupling the density and the polarization
become 
\be
n&=&\sum\limits_p f= {m_e\over 2 \pi \hbar^2} \left ( \epsilon_f+\epsilon_\beta\right )
\nonumber\\
s&=&\sum\limits_p g=-{m_e\over 2\pi \hbar^2} \sqrt{\epsilon_\beta(\epsilon_\beta+2 \epsilon_f)+\Sigma_n^2}
\label{ns}
\ee 
with the spin-orbit energy $\epsilon_\beta=m_e\beta^2$. Further we have
\be
B_g^2={4\pi \hbar^2\over m_e}p n^2 {\epsilon_\beta\over \Sigma_n^2}
\ee
with the polarization $p=s/n$. 

\begin{figure}[]
\includegraphics[width=4.3cm]{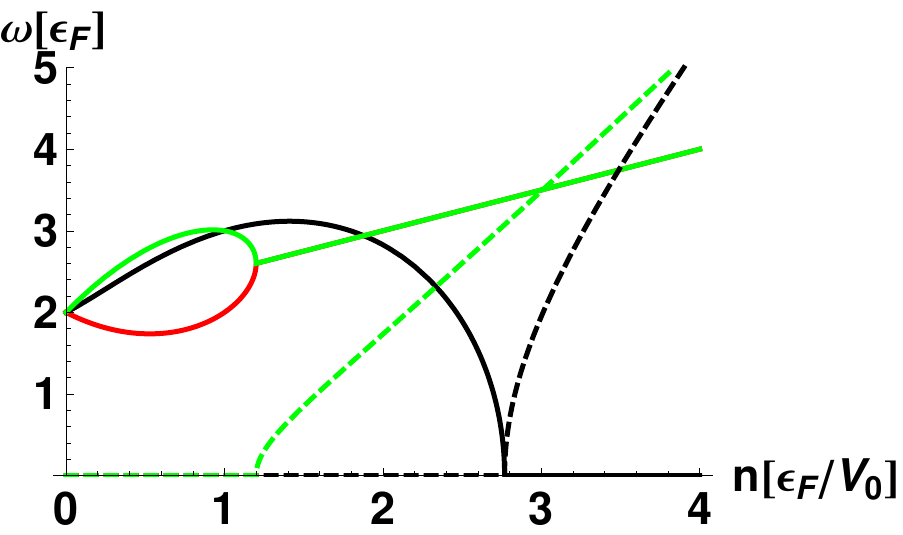}\includegraphics[width=4.3cm]{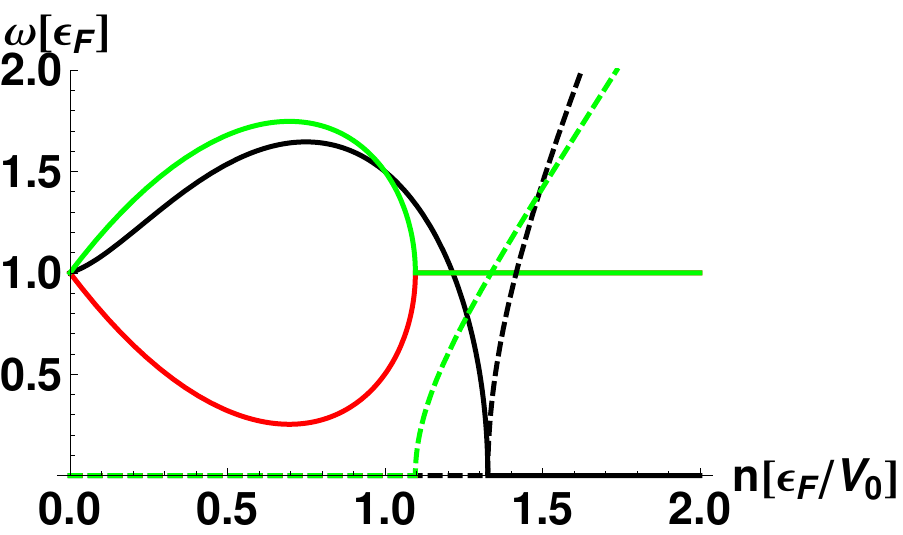}
\caption{\label{mode_T0_p05B1v1} The real (solid) and imaginary (dashed) part of $\omega$ as a function of the density for the effective magnetic field $\mu_B B^{\rm eff}=n V+\mu_B B=1\epsilon_f$, the polarization $p=0.5$ and the dimensionless potential $v_0=m_e V_0/2\pi \hbar^2=1$ at zero temperature. Left figure shows the selfconsistent result and right figure the non-selfconsistent one. Different branches of the single mode are distinguished additionally by different colors.}
\end{figure}

In Fig. \ref{mode_T0_p05B1v1} we plot the non-self-consistent and
self-consistent modes in dependence on the density. Using the polarization and
the scaling with the Fermi energy $\epsilon_f$ allows us to get rid of the spin-orbit energy. One sees that the selfconsistency leads to smaller modes at smaller densities. Up to a critical density, we do not have any imaginary part. As soon as the two spin modes vanish, a damping occurs that is symmetric in sign such that it denotes an instability. The third mode of (\ref{threemodes}) is vanishing at higher densities. 

\begin{figure}[]
\includegraphics[width=4.3cm]{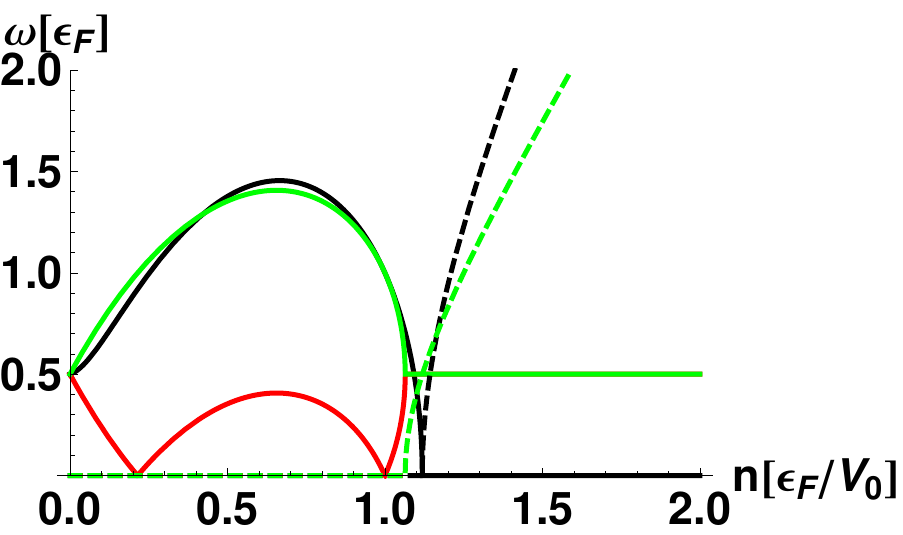}\includegraphics[width=4.3cm]{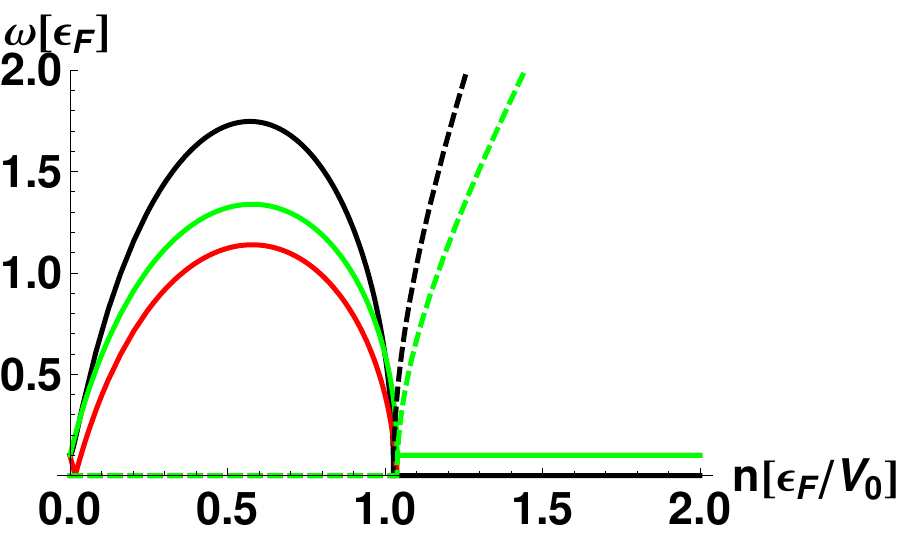}
\caption{\label{mode_T0_p05v1} The selfconsistent modes of figure \ref{mode_T0_p05B1v1} for effective magnetic fields $\mu_B B^{\rm eff}=n V+\mu_B B=0.5\epsilon_f$ (left) and $\mu_B B^{\rm eff}=n V+\mu_B B=0.1\epsilon_f$ (right).}
\end{figure}

The dependence on the effective magnetic field is seen in Fig. \ref{mode_T0_p05v1}, which shows that the two spin modes become nontrivial and vanish at the same density as the third mode for vanishing magnetic field.
  
The expansion of (\ref{threemodes}) in small spin-orbit coupling reads
\ba
\left (\omega_{\rm spin}+{i\over \tau}\right)^2&=
\left \{\begin{matrix} 4 \Sigma^2-8 \epsilon_f \epsilon_\beta
\cr
(4 \Sigma^2+8 \epsilon_f \epsilon_\beta)(1+{m_e\over 2 \pi \hbar^2}V_0)^2
\end{matrix}
\right .+o(\epsilon_\beta^2)
\end{align}
and the third mode $4 \Sigma^2-{8m_e V_0\over 2 \pi \hbar^2} \epsilon_f \epsilon_\beta$. This shows that the spin modes becomes
\ba
\omega+i/\tau&=\left \{\begin{matrix} i\left (2 \beta p_f-{\Sigma^2\over \beta p_f}\right )
\cr
\left (2 \beta p_f+{\Sigma^2\over \beta p_f}\right )\left (1+{m_e V_0\over 2 \pi \hbar^2}\right )
\end{matrix}
\right .+o(\epsilon_\beta^2,\Sigma^4)
\end{align}
 for small effective Zeeman fields
providing a linear dependence of the energy and damping on $\beta$. This is in
contrast to the influence of the Landau levels which provides quadratic dependencies \cite{BL04,E06}.

\subsection{Excitation with charged scattering}  

Now we consider the charged scattering with an impurity Coulomb potential
$V_0=e^2/\epsilon_oq^2$ or the scattering between charged particles as
meanfields and in the relaxation time approximation. Using the long-wavelength expansions of Appendix~\ref{expansion}, we obtain the following equation system from (\ref{final})
\be
&&\left  ({\omega_+^2\over \omega_p^2}+1-{i\omega_+\over \omega_p^2\tau}\right )\Psi_j=\V m\cdot \V \Psi_j
\nonumber\\
&&
2\sum\limits_p g(1-\V e\circ \V e)\V \Psi_j+ i \omega \V m\times \V \Psi_j=0
\label{modec}
\ee 
with $\V m=\V s_{q=0}=\V e_z s$. We used only the most divergent terms $\sim
1/q^2$ in the equation for $\delta \V s$ and have introduced the plasma
frequency $\omega_p^2=e^2 n/\epsilon_0 m_e$. The eigenmodes of (\ref{modec})
can be seen to decouple for density and spin modes since the equation provides
the eigenmode as a vanishing determinant of the matrix in front of $\V \Psi_j$. The density eigenmodes are given by the vanishing left side of the first equation which implies $\V m\perp \V \Psi_j$ which means that we have only transverse spin modes with respect to the effective magnetization axes $\V m$ of (\ref{s}). The frequency of the density modes is just the damped plasma oscillation,
\be
\omega_{n}=-{i\over 2 \tau}\pm \sqrt{\omega_p^2-{1\over 4\tau^2}}
\ee
and the spin oscillation becomes
\be
\omega_{\rm spin}=-{i\over \tau} \pm 2 [\Sigma+o(b^2)]
\ee
where we have used the linear expansion in $b$ of the last paragraph.
Compared with the spin mode of the neutral excitation (\ref{modes0}), we see that the term $s V_0$ is absent.

\subsubsection{Dielectric function}

\begin{figure}[]
\includegraphics[width=4.4cm]{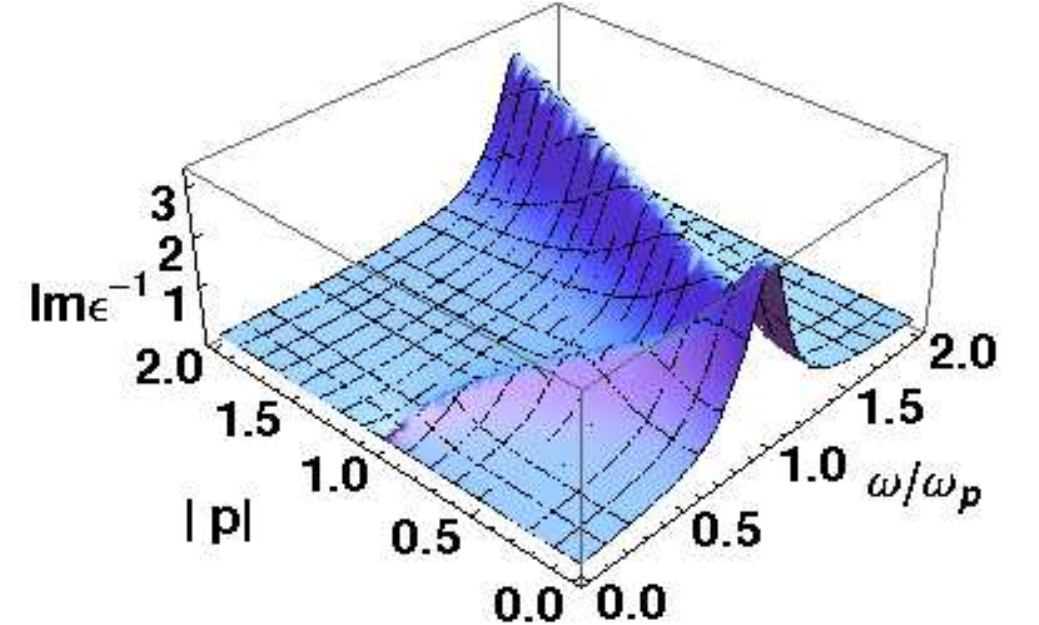}\includegraphics[width=4.4cm]{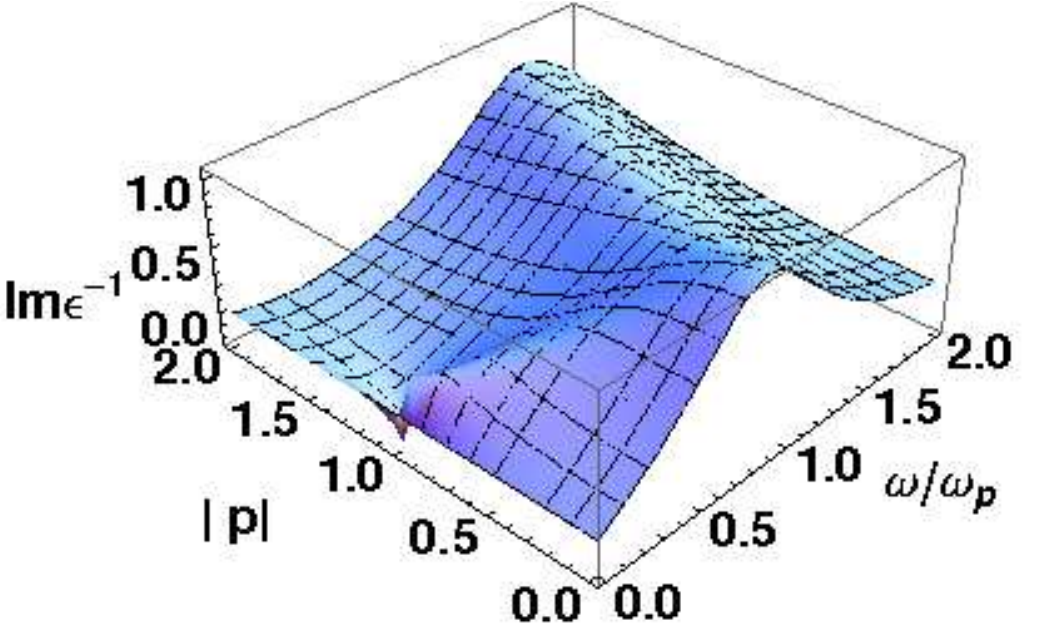}
\caption{\label{excit3D_nu0_3} The long-wave excitation function (\ref{eps0}) for a collision frequency $\tau^{-1}=0.3\omega_p$ (left) and $\tau^{-1}=1\omega_p$ (right).}
\end{figure}

The response function (\ref{res}) describes
the density change with respect to the external potential $\delta n=\chi
\Phi$ while the polarization function $\delta n=\Pi \Phi^{\rm ind}$ is the density variation with respect to
the induced potential $\Phi^ {\rm ind}V_q\delta n+\Phi$. Therefore one has
$\chi=\Pi/(1-V_q\Pi)$, and the dielectric function as a ratio of the induced
to the
external potential is
\be
{1\over \epsilon}=1+V_q \chi.
\ee
If we expand the response function up to quadratic orders of the wave vector, as performed in appendix~\ref{expansion},
the result for the dielectric function can be compactly written as
\be
\epsilon(\omega,q)=1-{1\over {1\over 1-\epsilon(\omega,0)}-{q^2\over
    \kappa^{\rm eff}(\omega)}}
\label{eps}
\ee
where the long-wavelength dielectric function can be expressed as
\ba
&\epsilon(\omega,0)=\epsilon_\omega \nonumber\\
&+p^2\left (1\!-\!{1\over \epsilon_\omega}\right )
\left [
1\!-\!\epsilon_\omega\!-\!{B_f^2\over \epsilon_\omega}
\left (
(1\!-\!\epsilon_\omega)^2\!-\!{\omega_p^2\over
    \omega^2(1\!+\!p^2)}
\right ) \right ]
\label{eps0}
\end{align}
in terms of the Drude's expression
\be
\epsilon_\omega=1-{\omega_p^2\over \omega (\omega+{i\over \tau})}
\label{epsmerm}
\ee
and the effective polarization 
\be
p={s\over n}={n_\uu-n_\dd\over n}-{B_g^2\over 2n}.
\label{poleff}
\ee 
Here we use the zero temperature result for linear spin-orbit coupling 
\be
B_g^2= {2 p \over 1+p^2} B_f^2
\ee
with (\ref{Bg}).
Let us discuss the long-wavelength limit of the dielectric
function first without spin-orbit coupling $B_f=0$ but finite polarization (\ref{poleff}). This corresponds to the treatments of two-fluid models, e.g. one-dimensional quantum wires \cite{Bala14}, with finite polarization. 
In figure \ref{excit3D_nu0_3} we plot the
excitation function which yields the weight of the collective modes as a
function of frequency and polarization. There we plotted a larger range
of polarizations. Since the latter one is an effective one according to
(\ref{poleff}) we can have values smaller than $-1$. A smaller relaxation time leads to a higher damping of modes, of course.

\begin{figure}[]
\includegraphics[width=8cm]{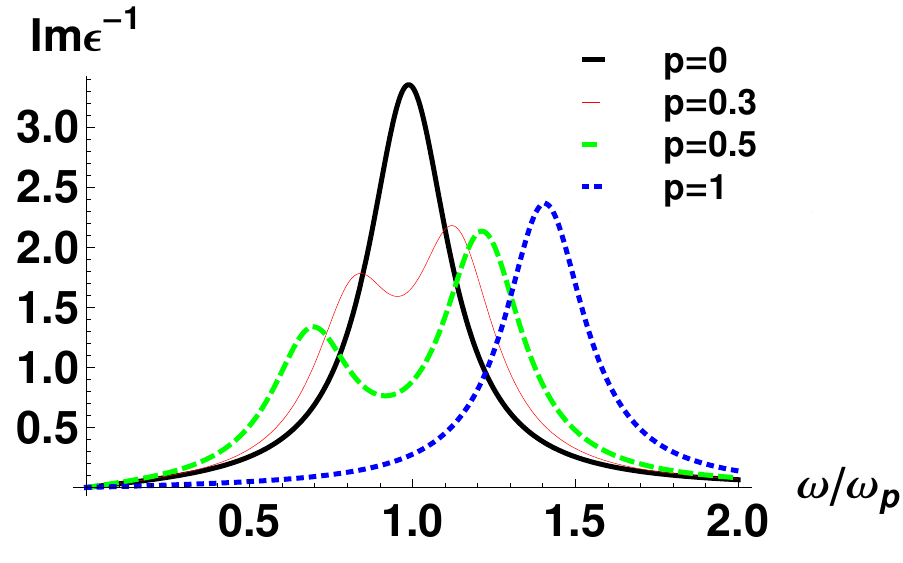}
\caption{\label{excit_m0_1w} The long-wave excitation function (\ref{eps0}) for different polarizations versus frequency as cuts of figure \ref{excit3D_nu0_3} for $\tau^{-1}=0.3\omega_p$.}
\end{figure}

We see the appearance of two
collective modes with increasing polarization. The plasma mode becomes
split in a fast decaying mode with increasing polarization and a mode which
becomes sharper again with increasing polarization. This is illustrated in
figure \ref{excit_m0_1w} as cuts for special polarizations.


\begin{figure}[]
\includegraphics[width=8cm]{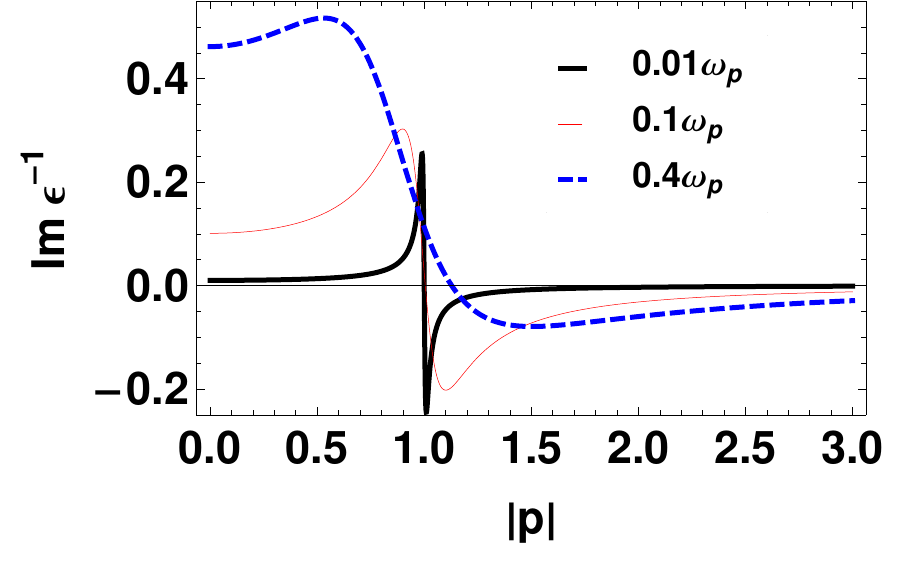}
\caption{\label{excit_w0_04m} The long-wave excitation function (\ref{eps0})
  for three small frequencies versus effective polarization (\ref{poleff}).}
\end{figure}

Near the point of vanishing first mode around $p=1$ the
excitation function becomes negative indicating an instability. This is
illustrated in the next figure \ref{excit_w0_04m}. The effective polarization as the sum of polarization and
spin-orbit coupling term $B_g$ can lead to negative excitation functions
indicating an instability. We will interpreted this instability as a de-mixing
of spin states later when discussing the screening length.


\begin{figure}[]
\includegraphics[width=8cm]{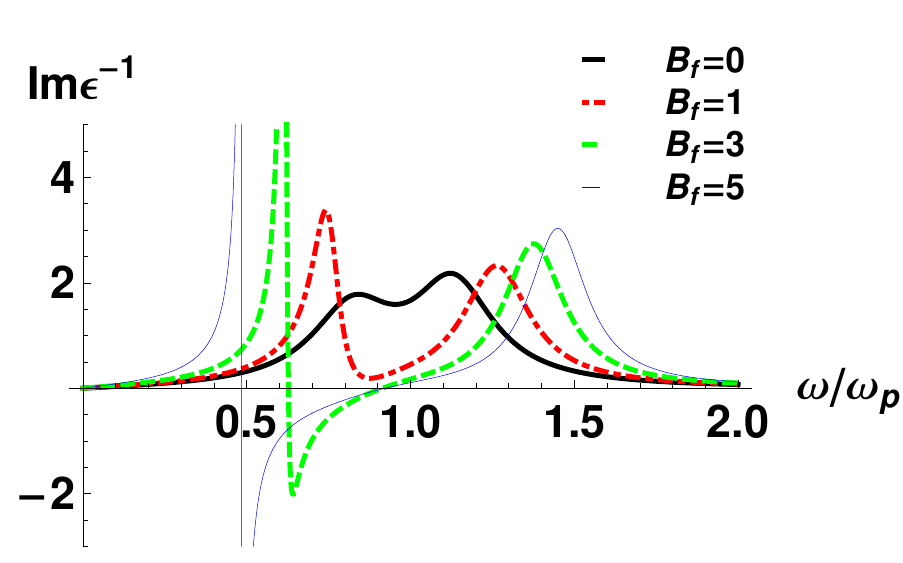}
\caption{\label{excit_Bf_w} The long-wave excitation function (\ref{eps0})
  for different spin-orbit couplings $B_f$ versus frequency with a fixed
  polarization $p=0.3$ of figure \ref{excit_m0_1w}.}
\end{figure}

\begin{figure}[]
\includegraphics[width=8cm]{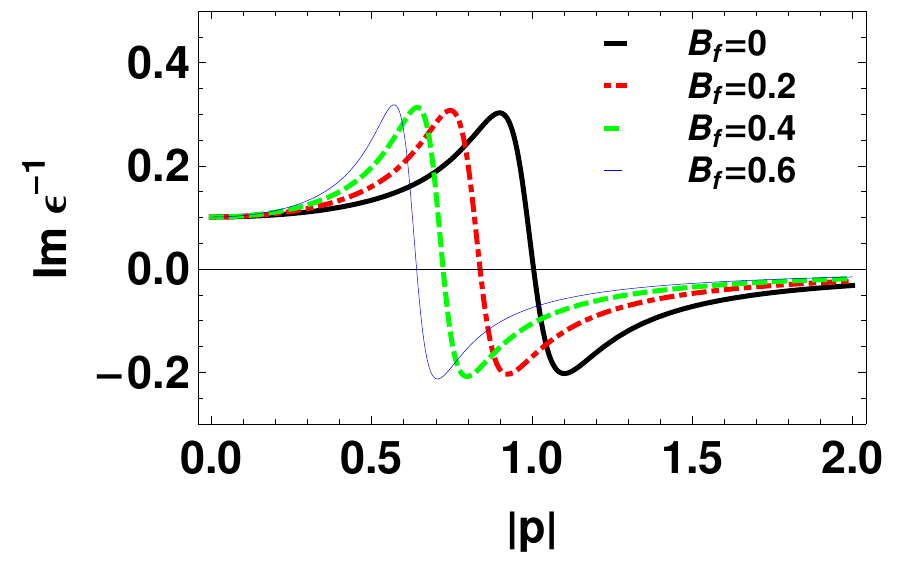}
\caption{\label{excit_Bf_w0_04} The long-wave excitation function (\ref{eps0})
  for different spin-orbit couplings $B_f$ versus effective polarization
  (\ref{poleff}) with a fixed
  polarization $|p|=1$ and frequency $\omega=0.1\omega_p$ of figure \ref{excit_w0_04m}.}
\end{figure}

Next we consider the influence of the spin-orbit coupling $B_f$ which is given
in figure \ref{excit_Bf_w} for different values and a fixed polarization. We
see that the spin-orbit coupling has basically the same effect as an
additional polarization. Above a certain spin-orbit coupling the excitation function becomes negative indicating an instability. This is also visible in the figure \ref{excit_Bf_w0_04}.

The range where the excitation function becomes negative indicating an
instability is plotted in the next figure \ref{excit_region}. 
This range indicates a spin domain separation and
becomes large for increasing spin-orbit coupling parameter $B_f^2$ of (\ref{Bg3}). We interpret this as spin segregation as observed in \cite{LHWC02} and described in \cite{WNCC02}.

\begin{figure}[]
\includegraphics[width=4.5cm]{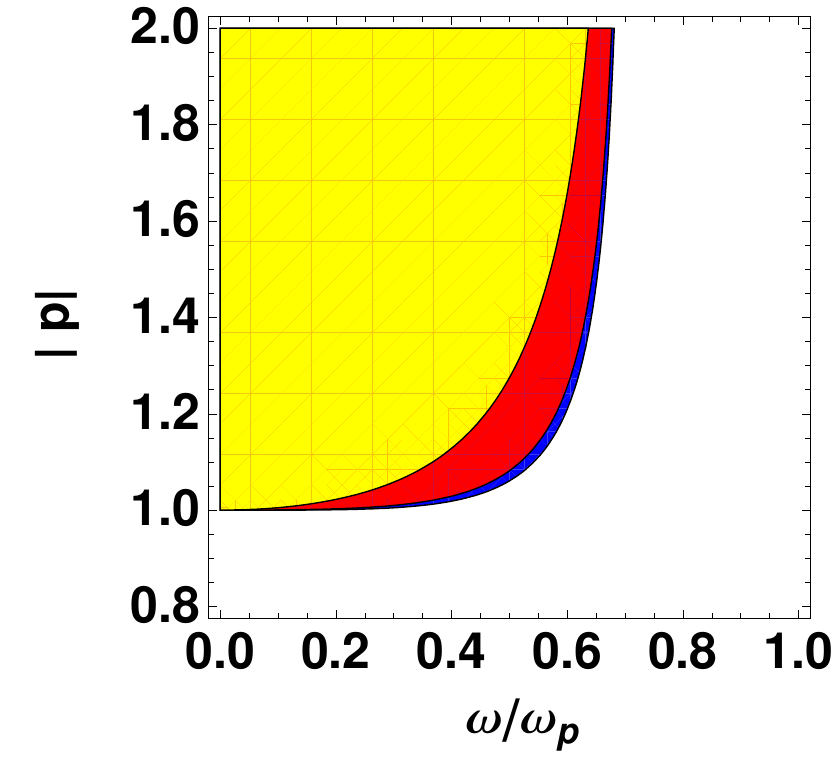}\includegraphics[width=4.1cm]{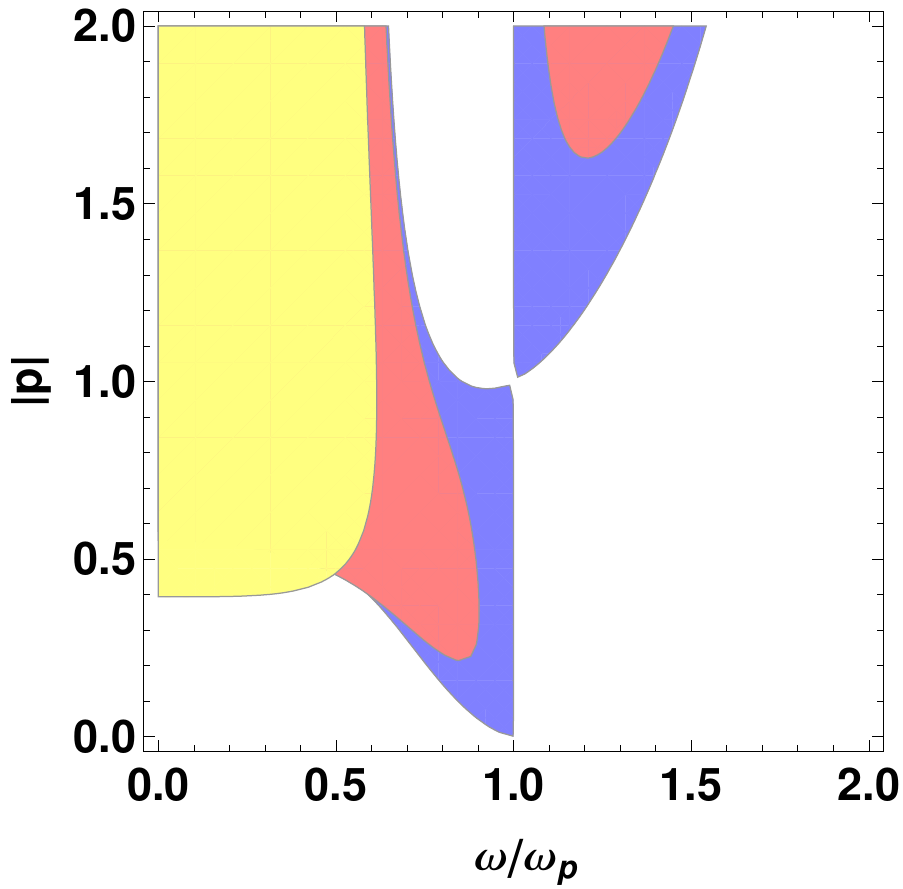}
\caption{\label{excit_region} The region where the excitation function becomes
  negative versus frequency and polarization
for $B_f=0$ (left) and $B_f=2$ (right)
. The upper range (yellow) is for $1/\tau=1\omega_p$, the middle
  (red) for
  $1/\tau=0.3\omega_p$ and the bottom (blue) one for $1/\tau=0.001\omega_p$.}
\end{figure}


\paragraph{Screening length}
Next we discuss the effective dynamical screening length $\chi^{\rm
  eff}(\omega)$ of (\ref{eps}) which can be expressed shortly in terms of (\ref{epsmerm}) as
\ba
&\left ({\kappa\over \kappa^{\rm eff}(\omega)}\right )^2={1\over 1\!-\!i \omega
  \tau}\!-\!p \kappa_V^2 {\epsilon_\omega\!-\!1\over \epsilon_\omega}
\left \{
1\!+\!{1\over (1\!-\!\epsilon_\omega)(1\!-\!q_\omega)}
\right .
\nonumber\\
&\left .
\!+B_f{\epsilon_\omega\!-\!1\over \epsilon_\omega(1\!-\!q_\omega)}
\left [
1\!+\!\left (1\!+\!{2(1\!-\!i\omega \tau)\over \omega_p^2 \tau^2(\epsilon_\omega(1+q_\omega)-1)}\right )
\right . \right .\nonumber\\
&\left . \left . \times
\left ({1\over
    (1\!-\!\epsilon_\omega)(1\!-\!q_\omega)}\!-\!q_\omega\right )
\right] 
\right \}
\label{kdyn}
\end{align}
and the short-hand notation $q_\omega=p^2(\epsilon_\omega-1)/\epsilon_\omega$.
The static limit where $(\epsilon_\omega-1)/\epsilon_\omega\to 1$ and
$\epsilon_\omega\to\infty$ reads therefore
\ba
\left ({\kappa\over \kappa^{\rm eff}(0)}\right )^2=1-p \, \kappa_V^2\left [1+B_f
\left (1-{2 p^2\over \omega_p^2 \tau^2 (1-p^4)} \right )\right ].
\label{keff0}
\end{align}
We have abbreviated $\kappa_V^2=(V+\mu_B B/n)\partial_\mu n=\mu_B B^{\rm eff}\partial_n n/n$. This result is explicitly an analytic long-wavelength expression of the influence of polarization on the screening length which was treated other wise by extensive numerics \cite{CKB79,S80}.

\begin{figure}[]
\includegraphics[width=8.5cm]{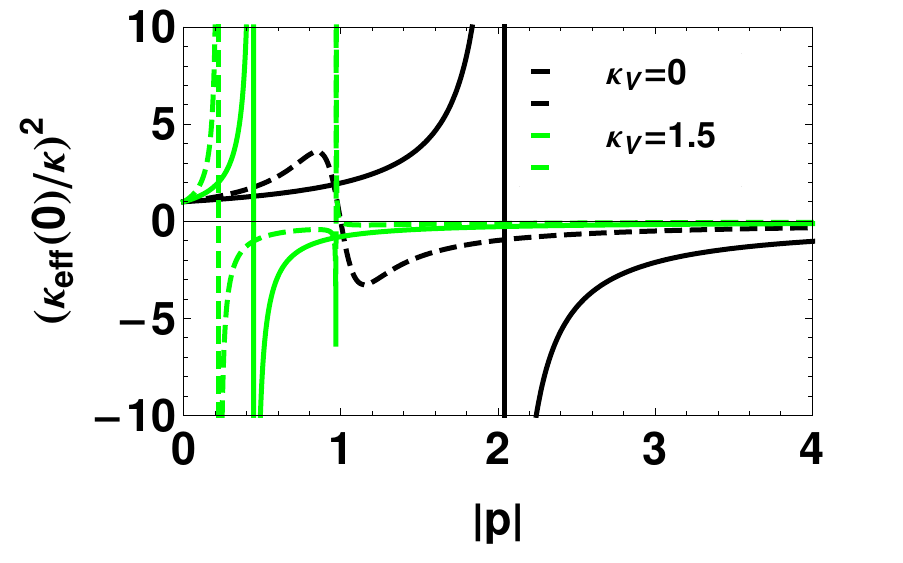}
\caption{\label{keff_p_V_0} The static effective screening length
  (\ref{keff0}) versus polarization for different $\kappa_V$ and $B_f=0$
  (solid) and $B_f=1$ (dashed).}
\end{figure}

The static
screening length (\ref{keff0}) changes only for finite $\kappa_V^2$ which means a finite
magnetic field or ferromagnetic impurity polarization $V$. In other words 
we need
a preferred direction of motion in order to see a change of static screening
length. In the latter case it is then dependent on the spin-orbit coupling
$B_f$ as illustrated in figure \ref{keff_p_V_0}. One sees that the screening
length increases with increasing polarization for $\kappa_V^2\ne 0$ and
diverges at the zero of (\ref{keff0}) which provides the critical $\kappa_V^2$
in terms of the polarization $p$ and the material parameter $\omega_p \tau$. 
This
instability appears here in the spatial screening length and we can interpret it as a spatial
domain separation of spin-polarized electrons known as domain wall formation \cite{NP12,CNB00}.
Interestingly, for finite spin-orbit coupling there appears an upper singularity
at $p=1$.

%
%

\begin{figure}[]
\includegraphics[width=4.2cm]{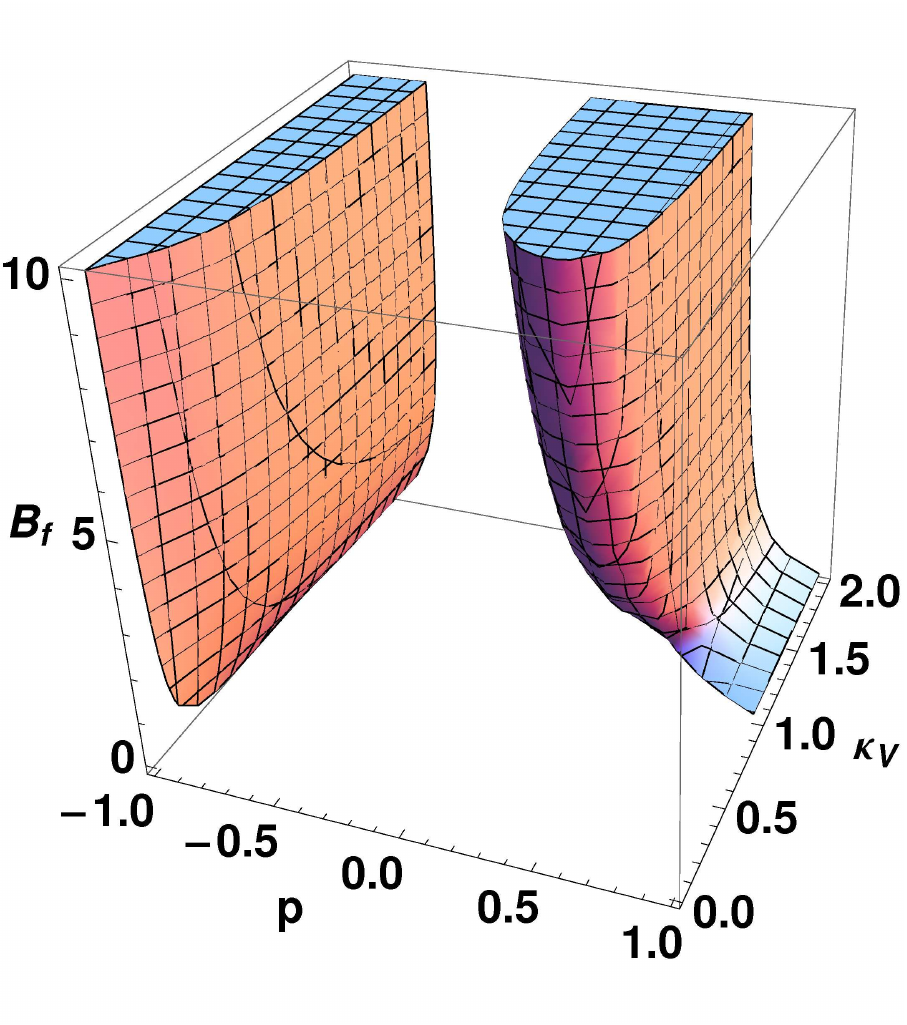}\includegraphics[width=4.2cm]{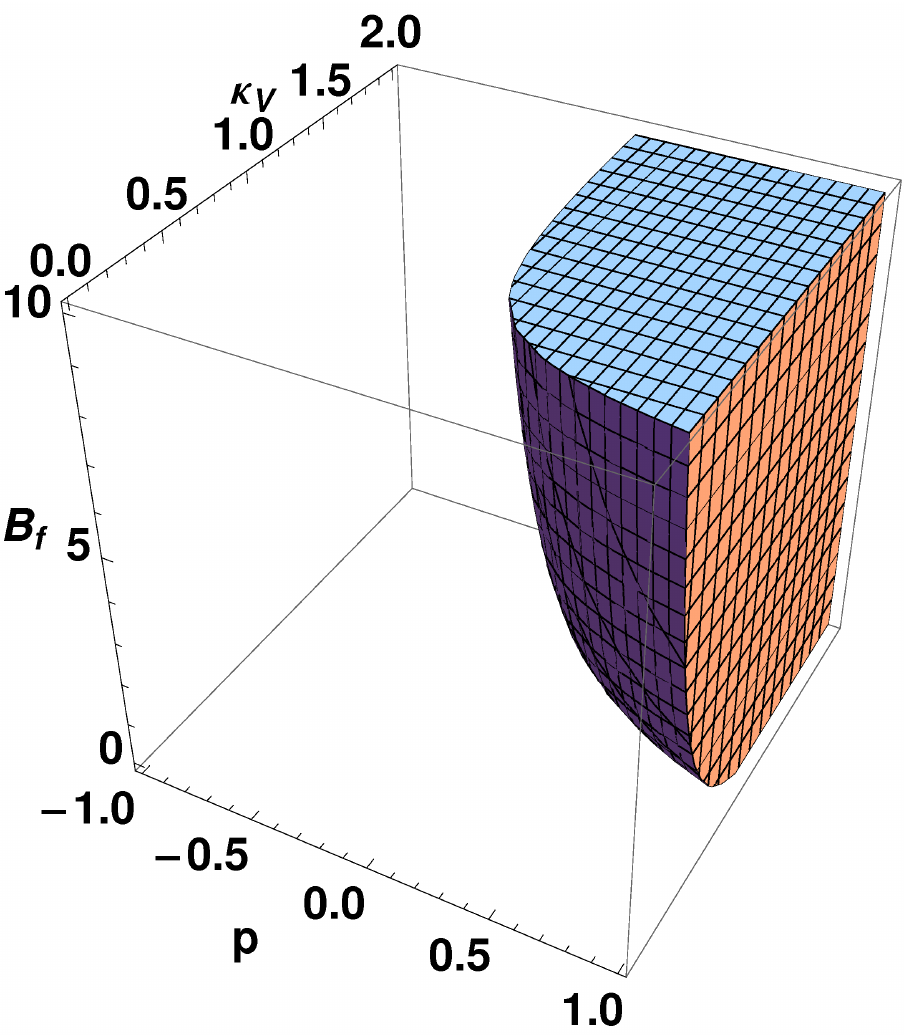}
\caption{\label{range_keff_stat_p_kv_Bf_n_1} (Color online) The range where the static effective screening length
  (\ref{keff0}) becomes negative for $1/\tau=1\omega_p$ (left) and $1/\tau=0\omega_p$ (right).}
\end{figure}

The range where the real part becomes negative is plotted in figure
\ref{range_keff_stat_p_kv_Bf_n_1}. With increasing collision frequency
the range of instability becomes smaller.

Comparing the case with vanishing collision frequency in figure
\ref{range_keff_stat_p_kv_Bf_n_1} we see that only one range at positive
polarization appears. In other words there appears an
asymmetric second range in the instability due to the collisions for positive and negative polarizations.


It is interesting to discuss the dynamical screening length as well. First one notes
that the correct static limit (\ref{keff0}) only appears if we have a
relaxation damping $1/\tau$ which drops out of the result. 
In contrast if we first set $1/\tau\to 0$ before the static limit we would obtain
\be
\lim\limits_{\tau \to\infty}\left ({\kappa^{\rm eff}(\omega)\over \kappa}\right )^2={(\omega^2-1)(\omega^2+p^2-1)\over \kappa_V^2 p (\omega^4+p^2-1)}
\ee
leading to the wrong static limit $-1/p\kappa_V^2$ which is clearly unphysical.
Therefore an even infinitesimal friction is necessary in order to ensure the correct static screening length. One can see this also from the limit of vanishing polarization $p\to 0$ which yields
\be
\lim\limits_{p\to 0}\left ({\kappa^{\rm eff}(\omega)\over \kappa}\right )^2=1-i\omega \tau.
\ee

\begin{figure}[]
\includegraphics[width=8cm]{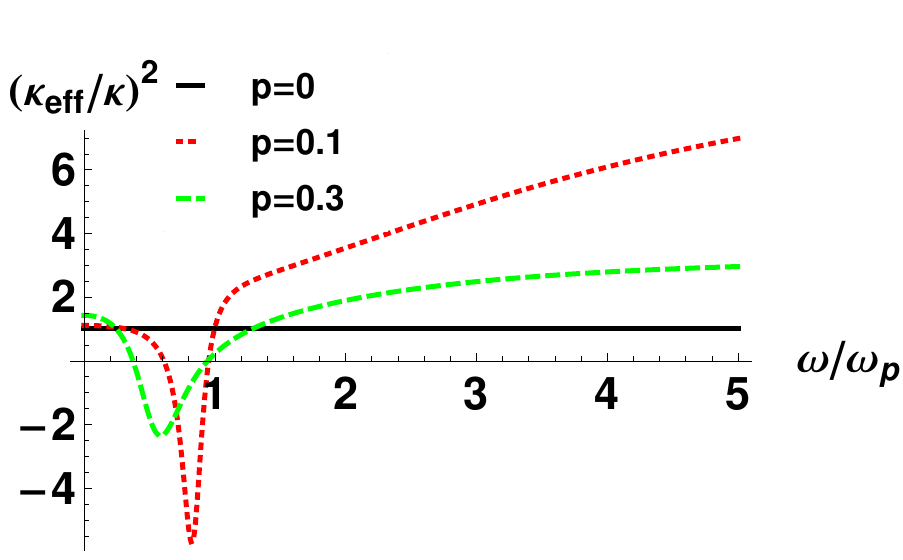}
\caption{\label{keff_w_p_0_03} The dynamical effective screening length
  (\ref{kdyn}) versus frequency for different polarizations, $\kappa_V=1$,
  $1/\tau=0.3\omega_p$ and $B_f=0$.}
\end{figure}

\begin{figure}[]
\includegraphics[width=8cm]{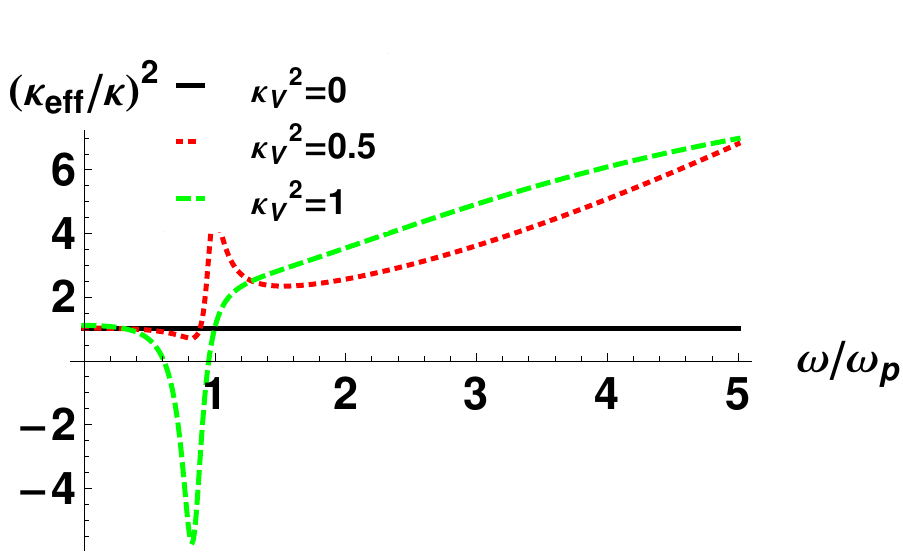}
\caption{\label{keff_w_kv_0_1} The dynamical effective screening length
  (\ref{kdyn}) versus frequency for different $\kappa_V$, a polarization
  $p=0.1$, $1/\tau=0.3\omega_p$, and $B_f=0$.}
\end{figure}

The dynamical screening length is plotted in figures \ref{keff_w_p_0_03} and
\ref{keff_w_kv_0_1} for  different cuts. Like in the static limit
above, at certain $\kappa_V$ the dynamical screening length becomes negative
indicating a domain-wall formation.

\begin{figure}[]
\includegraphics[width=4.4cm]{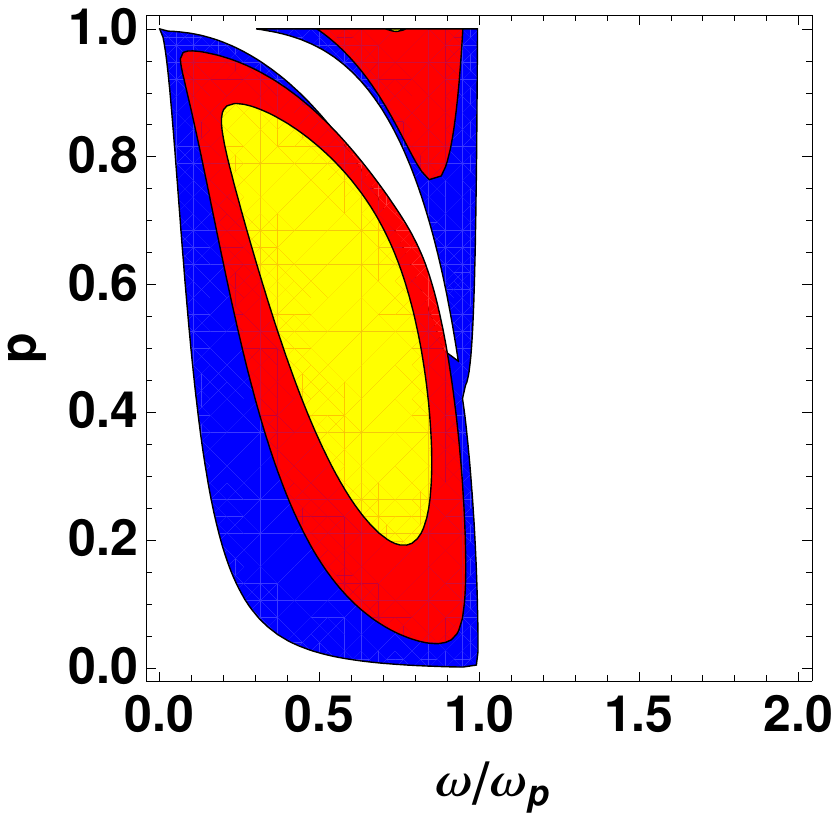}\includegraphics[width=4.4cm]{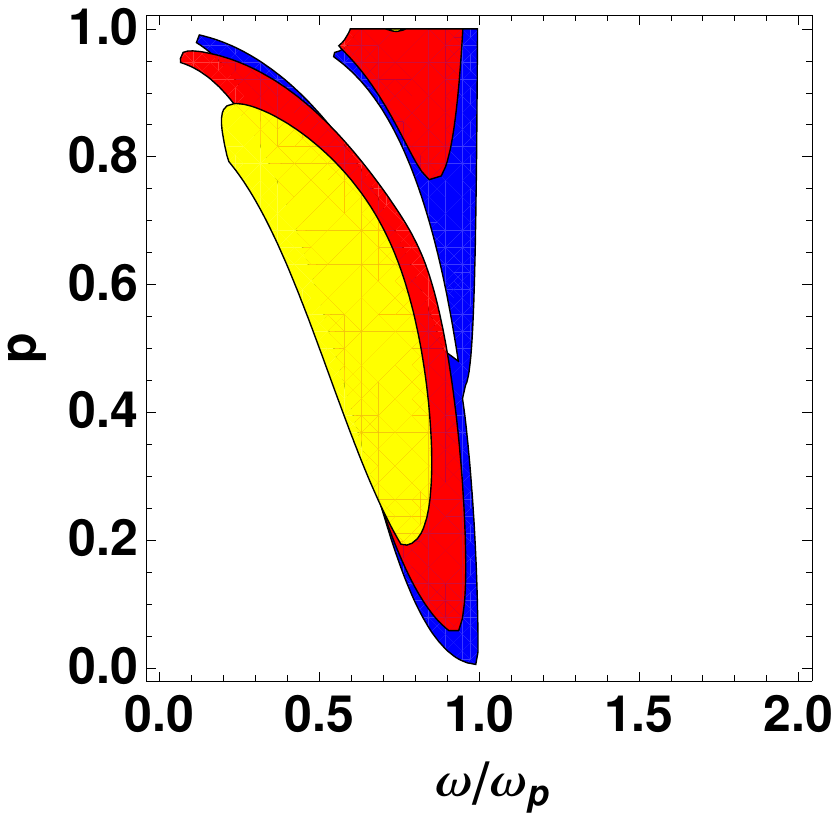}
\caption{\label{range_keff_w_p_kv_0} (Color online) Left: The range where the dynamical effective screening length
  (\ref{kdyn}) becomes negative versus frequency and polarization for
  $\kappa_V=1$. The inner range (yellow) is for $1/\tau=0.5\omega_p$, the middle
  (red) for
  $1/\tau=0.3\omega_p$ and the outer (blue) one for $1/\tau=0.1\omega_p$.
Right: additional demand that ${\rm Im} \kappa^{eff}>0$}
\end{figure}

The range where the real part becomes negative is given in figure
\ref{range_keff_w_p_kv_0}. With increasing collision frequency
the range of instability becomes smaller.
If we additionally demand that the imaginary part of the dynamical screening
length should be positive which means spatially unstable modes, we get a
smaller region. Above a certain collisional damping there is no such region.

%

\subsection{Spin response}

The spin response $\delta \V s=\V \chi_s \Phi$  can be calculated as well and we obtain with $\V q \Phi=i e \V E_q$
\be
{\delta \V s_q\over E}=i{e\over q} \V \chi_s={e n\over \omega_p}\begin{pmatrix}
s_1(\omega){b_1(q)\over q}+B_cs_2(\omega) {b_2(q)\over q}
\cr
s_1(\omega){b_2(q)\over q}-B_c s_2(\omega) {b_1(q)\over q}
\cr 
i {q \epsilon_0 \omega_p\over n e^2} s_3(\omega)
\end{pmatrix}
\label{ds}
\ee
which means we have an induced spin due to an applied electric field as used in microwave spectroscopy \cite{LLU13}. This is purely transverse to the $z$-direction of the effective magnetic and ferromagnetization field for long-wavelength. We consider this as the response of spin-Hall effect, described by the spin-orbit coupling $\V b(q)=(B_{\perp 1},B_{\perp 2},B_{\perp 3})$ according to (\ref{bsenk}) and (\ref{Bg3}).
Since $\V q || \V E$ this q-dependent spin-orbit coupling describes the
excitation due an external electric field. Oscillating electric currents are
used experimentally to create an effective magnetic field and ferromagnetic resonances \cite{FKWVZCCGJF15}.

The external magnetic field enters (\ref{ds}) by the dimensionless quantity
\be
B_c={\omega_c \epsilon_f\over n D \hbar \omega_p^2}=\epsilon_0 {B\over n D e \hbar} {\epsilon_f\over n}.
\label{Bc}
\ee  
The frequency-dependent functions (\ref{ds}) can be recast into the form
\be
s_1+1&=&-{1+p\over2} I(p)-{1-p\over 2}I(-p)\nonumber\\
s_2&=&{2 I_1(p)\over 1-p^2} -{I(p)-I(-p)\over p}+{I_2(p)\over p(1+p)}-{I_2(-p)\over p(1-p)} \nonumber\\
s_3&=&-{1+p\over2} I(p)+{1-p\over 2}I(-p)
\ee
with
\ba
I(p)&={1\over -1-p+{i\over \omega_p^2 \tau} \omega+{\omega^2\over \omega_p^2}}
&\to& -\omega_p {\sin{\gamma t}\over \gamma}{\rm e}^{-{t\over 2\tau}}
\nonumber\\
I_1(p)&={1\over 1-i\omega \tau}
&\to& {1\over \tau} {\rm e}^{-{t\over \tau}}
\nonumber\\
I_2(p)&={i \omega\over \tau \omega_p^2} I_1(p)
&\to& {1\over \tau}{\partial\over \partial t} {\sin{\gamma t}\over \gamma}{\rm e}^{-{t\over 2\tau}}
\end{align}
and 
\be
\gamma=\sqrt{(1+p)\omega_p^2-{1\over 4 \tau^2}}.
\label{frequenz}
\ee
Since we apply a frequency-constant electric field it means we have an instant disturbance of the system at time $t=0$ in the form $E(t)=E \delta(t)$. This field itself has to be subtracted from the response which is represented by the constant $s_1(\omega)+1$. Further we present the linearized result with respect to the spin-orbit coupling. A nonlinear analytic result with some more drastic simplifications can be found in \cite{SDGR06}.

The collisions are responsible for the damping of this oscillatory motion. Dependent on the temperature we will have a transition from collision-dominated damped motion towards an oscillatory regime as observed in \cite{BMZMHSSRS02}. This transition is here explicitly seen in the expression for $\gamma$ in (\ref{frequenz}) which turns the oscillatory behavior into an exponential one if 
\be
(1+p)\omega_p^2<{1\over 4 \tau^2}
\ee
which provides density, polarization, and (due to the relaxation time)
temperature-dependent criteria for such a transition.

It is now interesting to inspect the spin response for linear Dresselhaus and Rashba spin-orbit coupling.
If the electric field is excited in $x$-direction we have for Dresselhaus $b_1=\beta_D q_x/\Sigma_n, b_2=0$ and for Rashba $b_1=0,b_2=-\beta_R q_x/\Sigma_n$. This translates into the spin response
\be
\left . {\delta \V s_q\over E}\right |_D&=&{\beta_R q_x e n\over q\Sigma_n \omega_p} (s_1,-s_2 B_c,0) 
\nonumber\\
\left . {\delta \V s_q\over E}\right |_R&=&{\beta_D q_x e n\over q\Sigma_n \omega_p} (-s_2 B_c,-s_1,0)
\label{resspin}
\ee

If the electric field is excited in $y$-direction we have for Dresselhaus $b_2=-\beta_D q_y/q, b_1=0$ and for Rashba $b_2=0,b_1=\beta_R q_y/q$ and the response in (\ref{resspin}) are interchanged between Dresselhaus and Rashba. Therefore it is sufficient to discuss one of the cases, say excitation in $x$-direction. 
Let us concentrate on the Dresselhaus relaxation.
From (\ref{resspin}) we see that the external magnetic field $B_c$ causes ellipsoid trajectories. If it is absent, we have a mere linear-polarized damped oscillation in x-direction. 

In the next figure \ref{spintrajD_p_0_wc_1} we plot the trajectories for different polarizations.

\begin{figure}[]
\includegraphics[width=4cm]{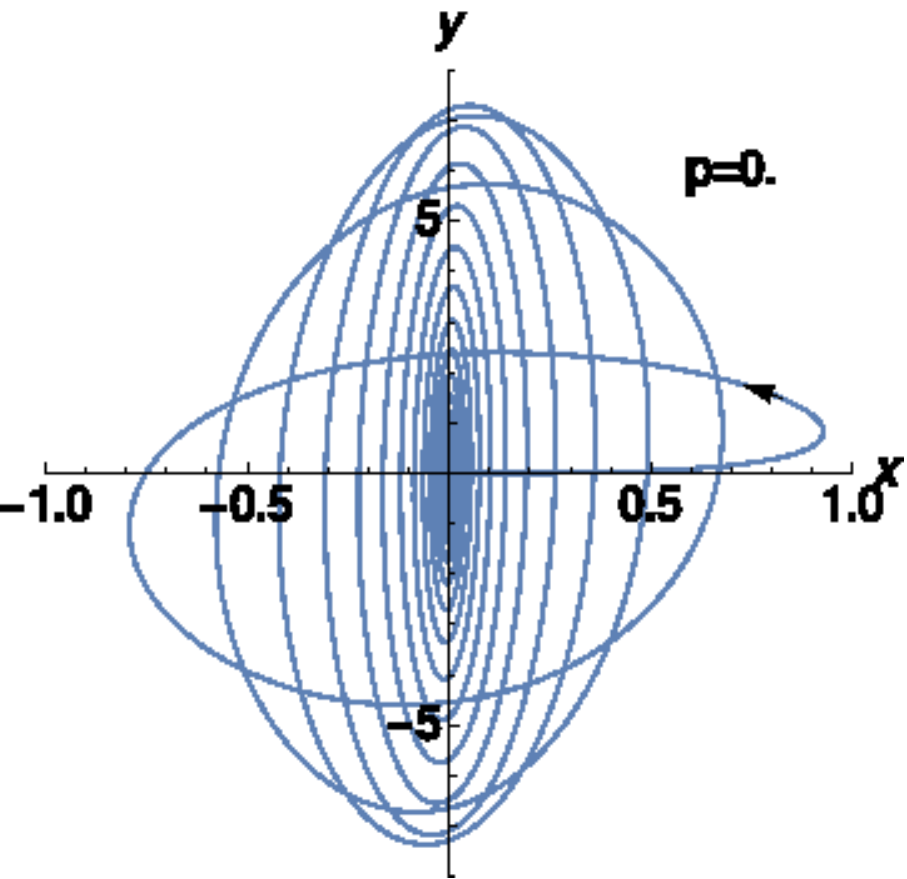}
\includegraphics[width=4cm]{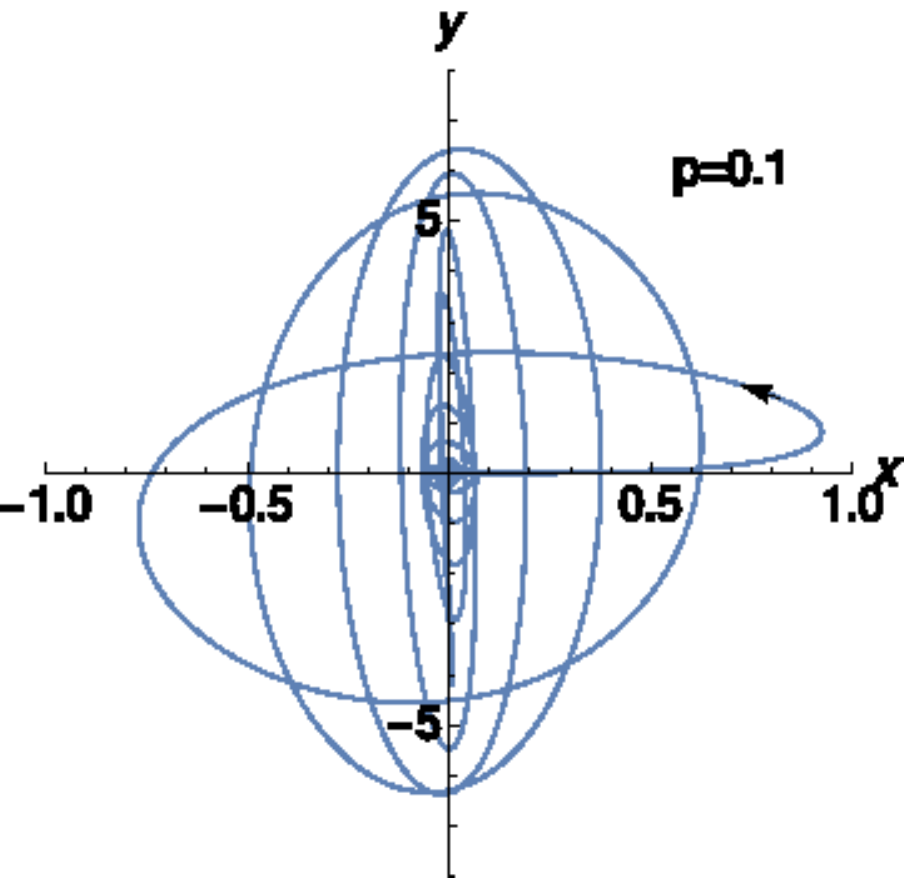}
\includegraphics[width=4cm]{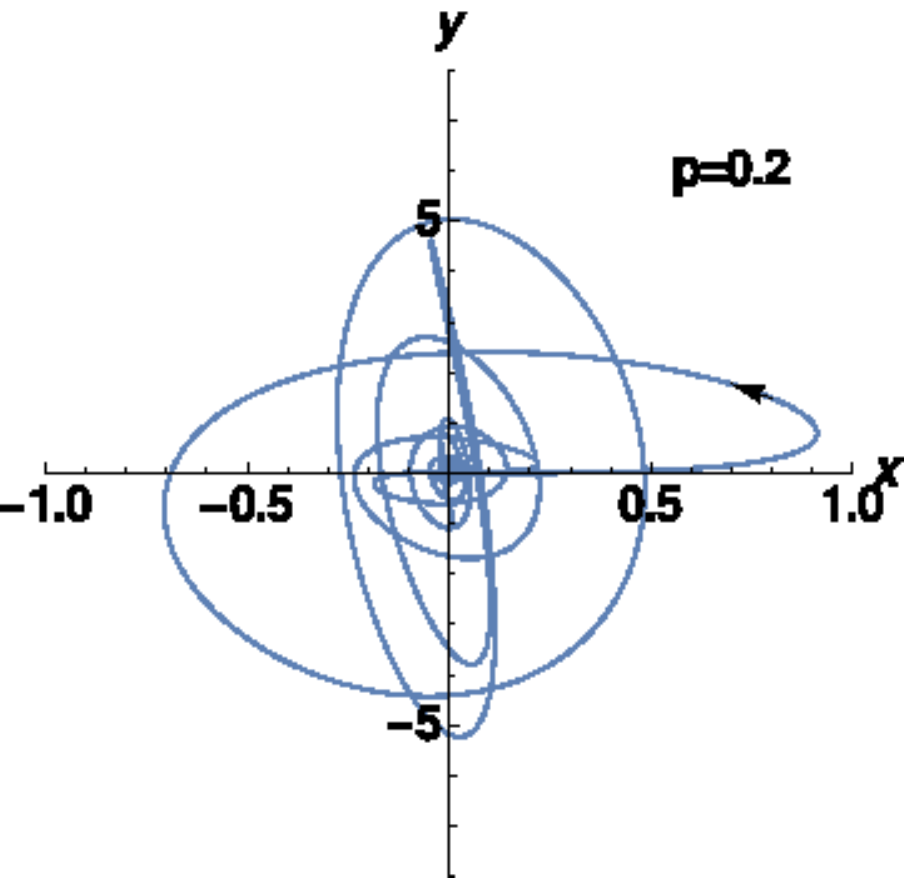}
\includegraphics[width=4cm]{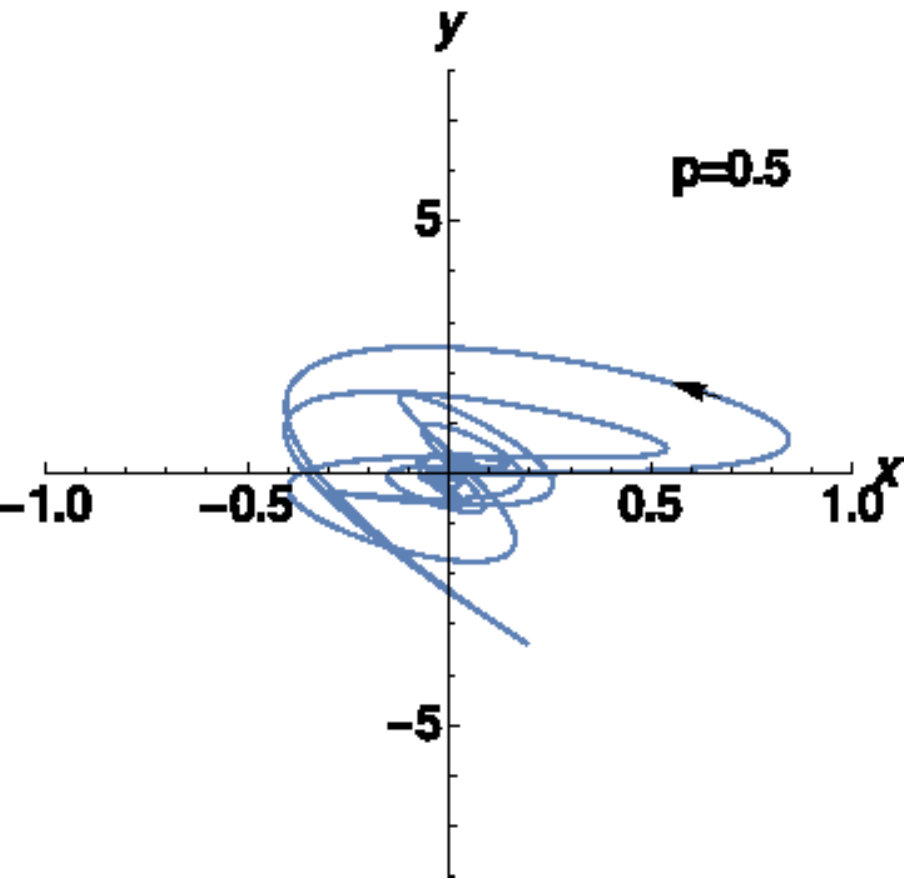}
\includegraphics[width=4cm]{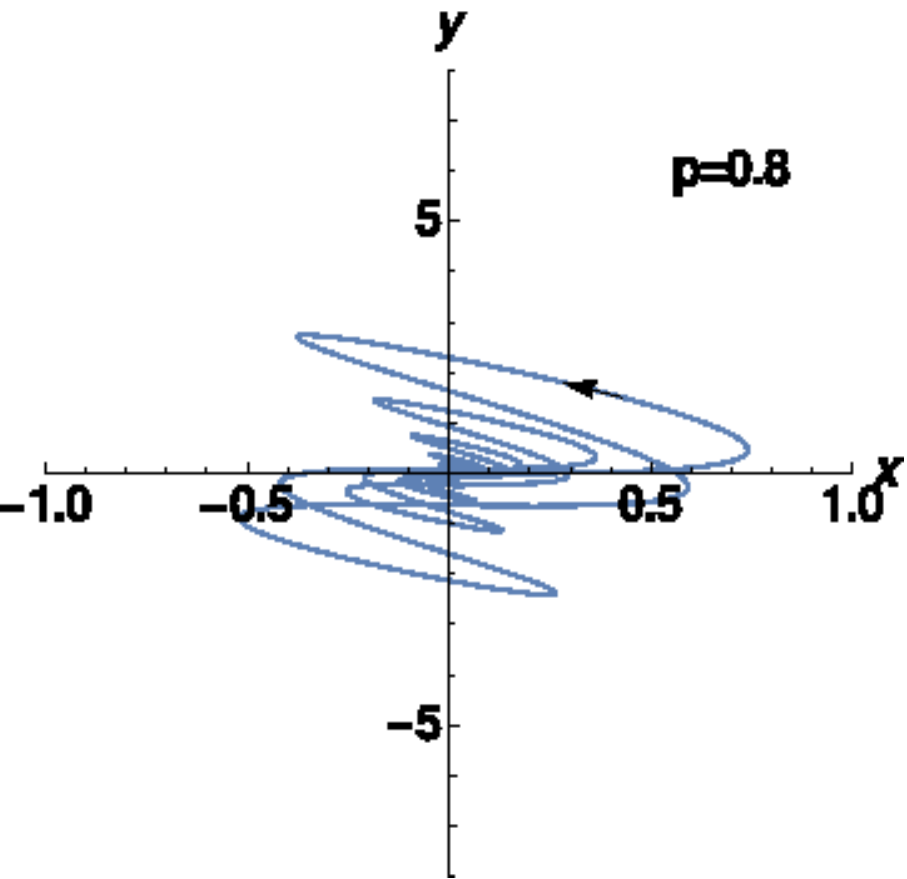}
\includegraphics[width=4cm]{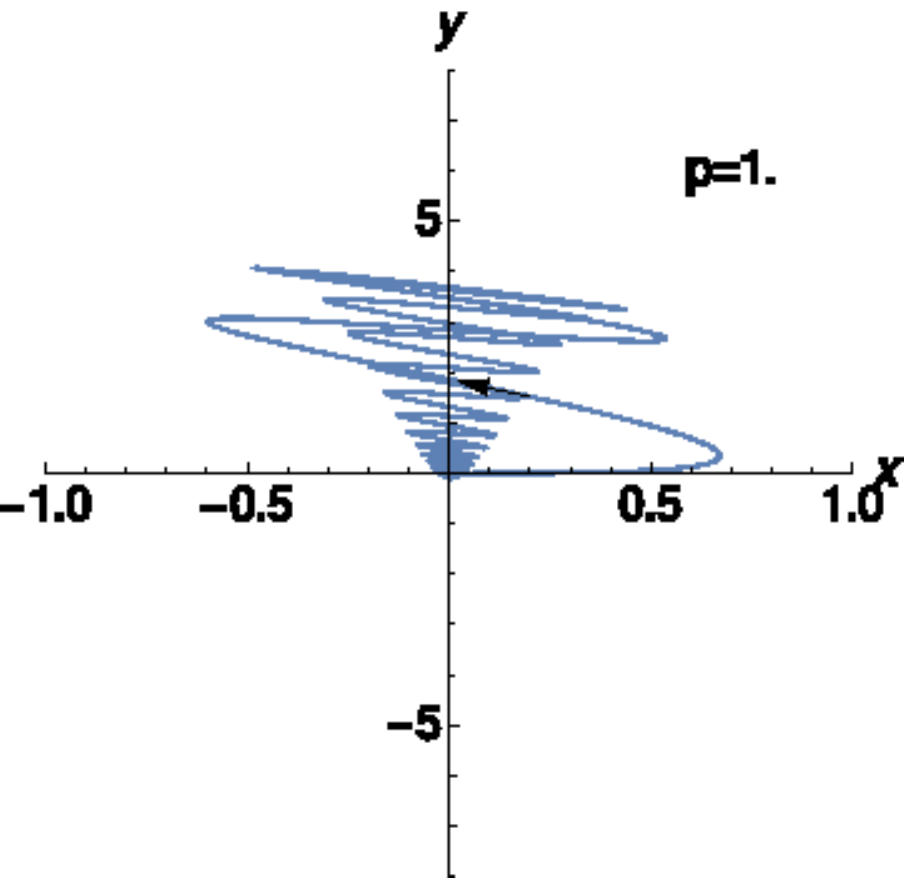}
\caption{\label{spintrajD_p_0_wc_1} (Movie online) The time-dependent trajectories of the induced spin with the disturbance of the electric field $\V E\delta(t)$ in $x$-direction for Dresselhaus spin-orbit coupling and
  $1/\tau=0.1\omega_p$. The external magnetic field was chosen $B_c=1$ according to (\ref{Bc}).}
\end{figure}

One recognizes that with increasing polarization the spin response turns to the perpendicular direction of the applied electric field which is a spin-Hall effect. Here we can see how the evolution of trajectories changes with increasing polarization.
The Rashba spin-orbit coupling will lead to the same curves but with $90^o$ clockwise rotation as one sees from (\ref{resspin}) too.

Since the change of spins (\ref{ds}) is different in each spatial direction triggered by spin-orbit coupling and split further by the magnetic field one can predict that this will lead to an anomalous spin segregation as was observed in \cite{DLCT08} and investigated in one-dimensional systems in \cite{EEL11}.


%
%

\begin{figure}[]
\includegraphics[width=4cm]{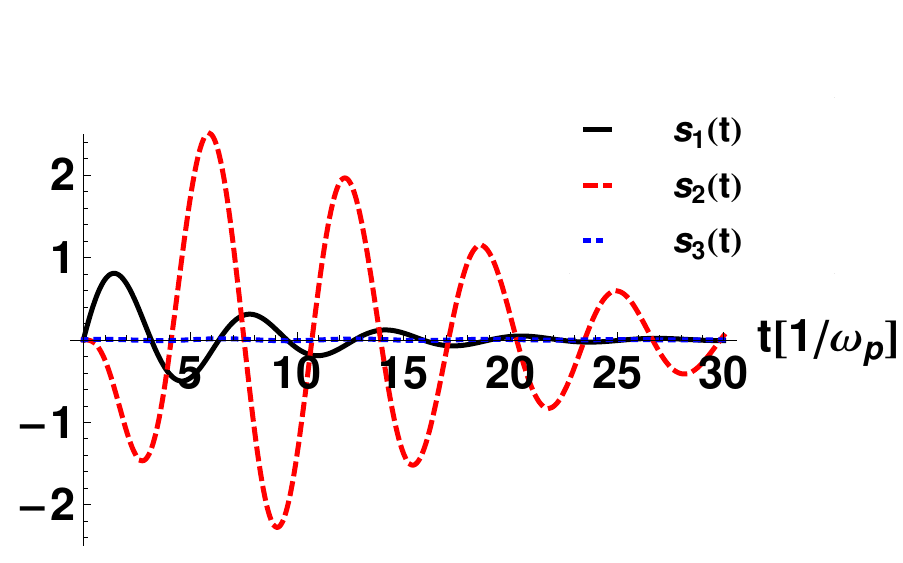}
\includegraphics[width=4cm]{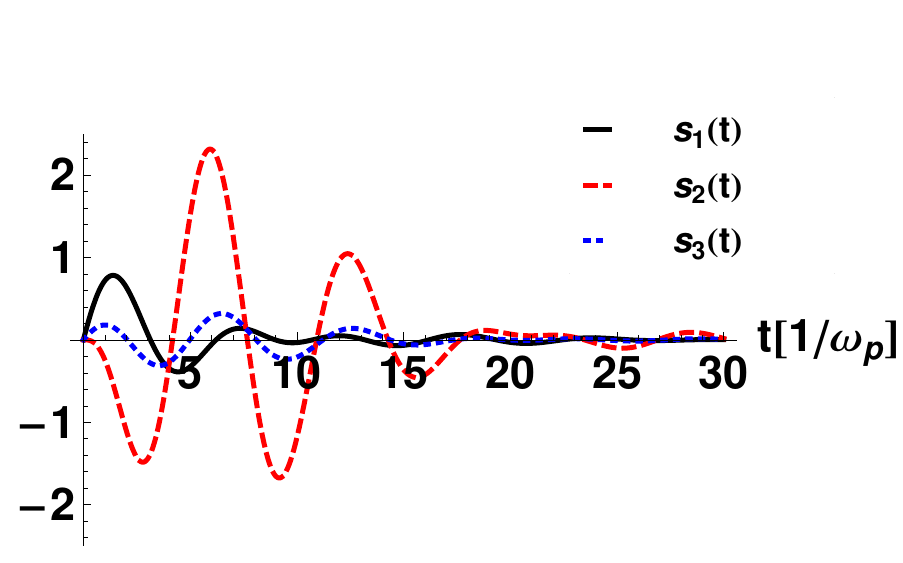}\\
\includegraphics[width=4cm]{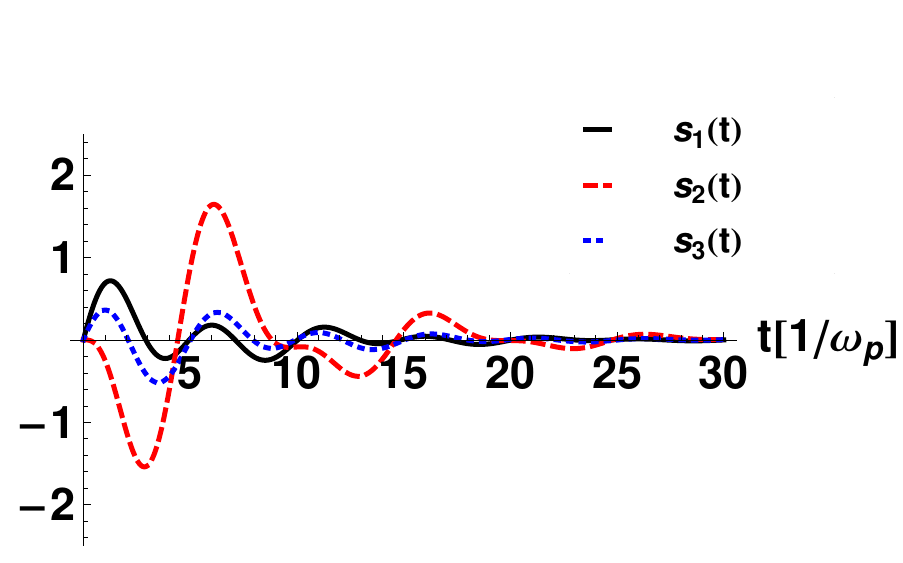}
\includegraphics[width=4cm]{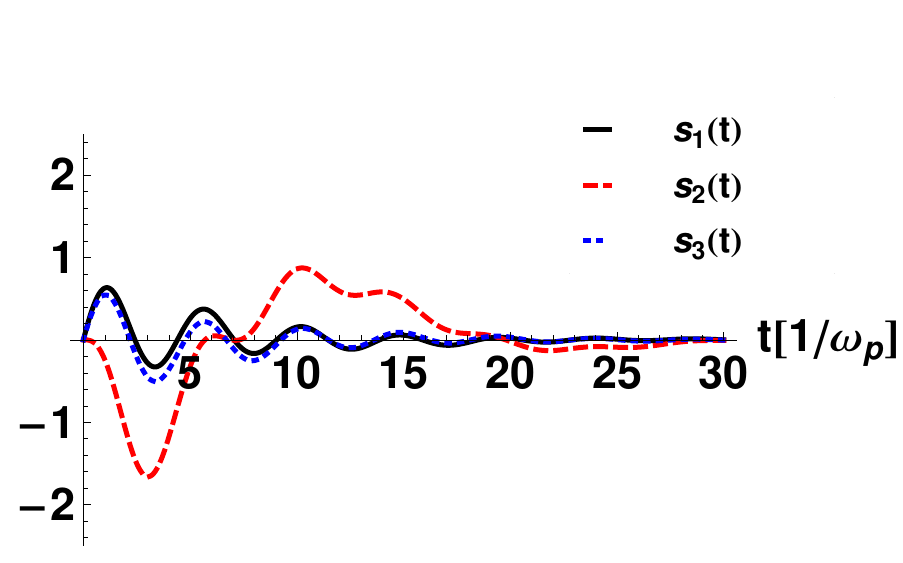}
\caption{\label{s123_nu} The time dependent spin response function (\ref{ds}) for $1/\tau=0.3\omega_p$ and polarizations $p=0.01, 0.3, 0.6, 0.9$ from upper left to lower right.}
\end{figure}

The spin dephasing time is of special interest \cite{MBGHRS00,SK02,WW03} where
one has found discrepancies between the experimental values and earlier
treatments. It is now quite difficult to extract dephasing times since the
envelope of the oscillation in each direction shows maxima and a quite nonlinear behavior as illustrated in figure \ref{s123_nu} by the constituent time-dependent functions of (\ref{ds}). One sees that besides oscillations with the frequency (\ref{frequenz}) the $s_1(t)$ possesses a maximum and all functions become quite nonlinear for higher polarizations. These components mix additionally due to the spin-orbit coupling and the magnetic field. If we, nevertheless, fit these time dependence to a damped exponential oscillator, we can extract the spin-dephasing time $\tau_s$ analogously to \cite{WW03}.  

\begin{figure}[]
\includegraphics[width=4.3cm]{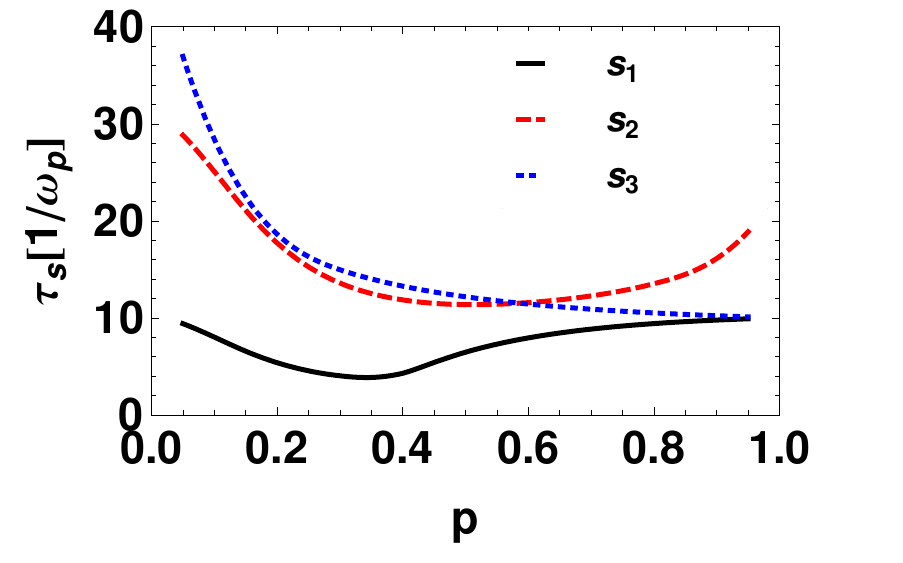}
\includegraphics[width=4.cm]{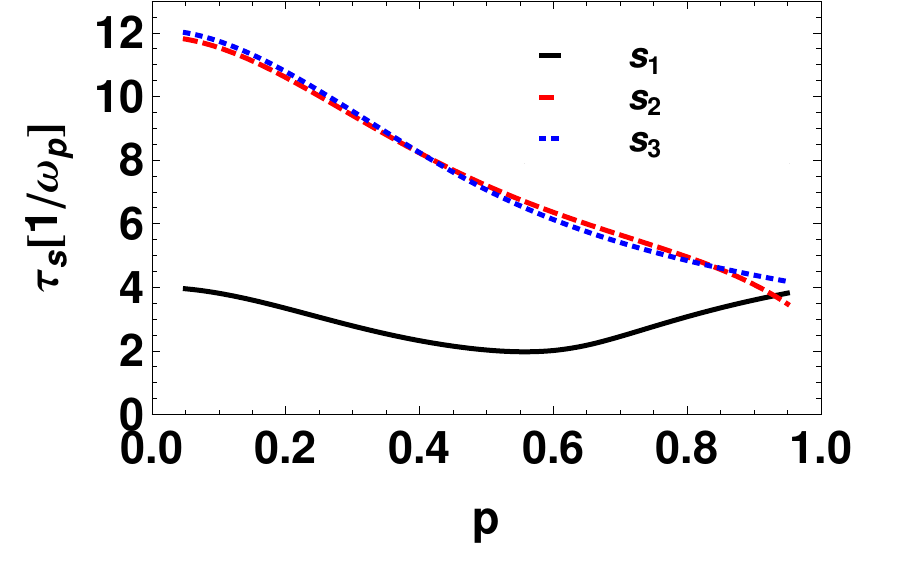}\\
\includegraphics[width=4.2cm]{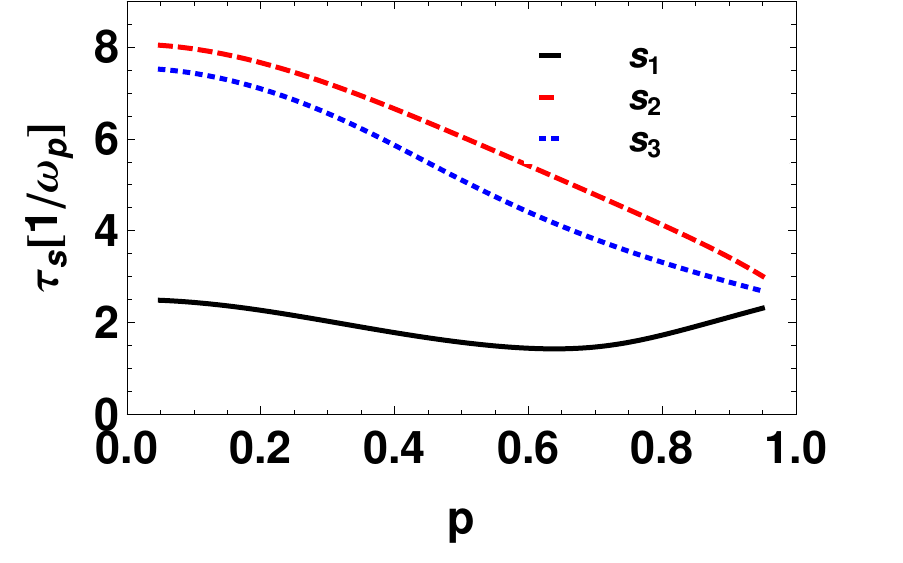}
\includegraphics[width=4.3cm]{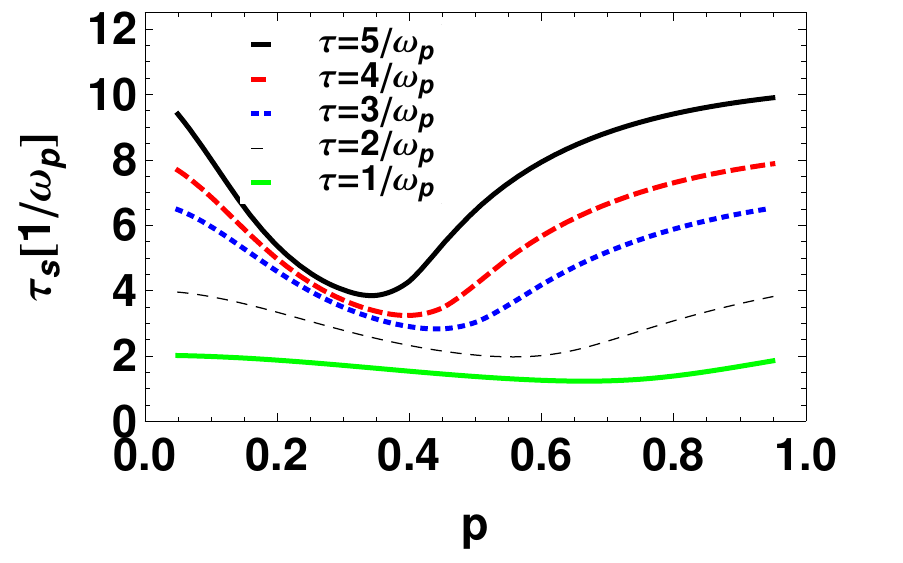}
\caption{\label{s123_dephas} The spin dephasing times (\ref{ds}) from exponential envelops for the different directions versus polarization with $1/\tau=0.2,0.5,0.8$ from upper left to lower left and $s_1$ for different relaxation times (lower right).}
\end{figure}

The results are given in figure \ref{s123_dephas}. The overall observation is
that the spin dephasing time is an order of magnitude larger than the
relaxation time. One sees that the $s_1$ component, which corresponds to the
$x$-component for the Dresselhaus and $y$ component for the Rashba coupling
has a minimum at a polarization which increases with increasing relaxation
time. The minima in $s_2$ and $s_3$ are not so pronounced and shift to larger
polarizations as well with increasing relaxation time. This result is
different from \cite{WW03} where only an increasing spin dephasing time in
dependence on the polarization has been reported. The combined effect of the
Rashba and the Dresselhaus coupling as well as the magnetic field mixes these results according to (\ref{ds}).

Here, we extracted the spin-dephasing time as an envelope of the precessional motion of the spins after a sudden distortion by an electric field. This is the Dyakonov-Perel mechanism of relaxation, e.g. investigated experimentally and theoretically in \cite{LFHGIRFSH07,LHHSFR07,LHHSFR08}

\section{Response with magnetic fields}

\subsection{Linearizing kinetic equation with magnetic field}
We want to consider the spin and density response to an external perturbing
electric field now under a constant bias of magnetic field. 
The magnetic field consists of a constant and an induced part $B(x,t)=B+\delta
B(x,t)$. Since the external electric field perturbation is produced due to an
external potential $U^{ext}$ one sees from the Maxwell equation that $\delta \dot {\V B}=-\nabla\times \delta \V E =0$, which means that all terms linear in the induced magnetic field vanish. 
It is convenient to work in velocity variables instead of momentum defined according to (\ref{vel}) and we use for the quasiparticle energy $\epsilon_p=p^2/2m_e+\Sigma_0$. We obtain finally from (\ref{kinet}) with the Larmor frequency $\omega_c={e B\over m_e}$
\begin{align}
&\left [-i\omega+iqv+{1\over \tau}+\left (v\times \omega_c\right )\p v\right ] \delta f
\nonumber\\&
+\left [{iq \over m_e}\p v\Sigma_{i}+(\V \tau^{-1})_i-\left ({\p v \Sigma_i \over m_e}\times \omega_c\right )\p v\right ] \delta g_{i}=S_0
\label{lina}
\end{align}
and 
\begin{align}
&\left [-i\omega+iqv+{1\over \tau}+\left (v\times \omega_c \right )\p v\right ] \delta g_{i}-2m_e(\Sigma\times \delta g)_i\nonumber\\&
+\left [{iq\over m_e}\p v\Sigma_{i}+(\V \tau^{-1})_i-\left ({\p v \Sigma_i\over m_e}\times \omega_c\right )\p v\right ] \delta f=S_i
\label{linb}
\end{align}
with the source terms arising from the external field $\V q \Phi=i e \V E$ and the induced meanfield variations (\ref{deltaS})
\ba
&S_0=
{iq\p v f\over m_e} \Phi+{\delta n\over \tau\partial_\mu n}\partial_\mu f_0
\nonumber\\
&\!+\!
\left (\!i{q \over m_e}\!+\!\omega_c\!\times\! {\p v^\Sigma \over m_e}\right )\!
\delta \Sigma_0\p v f\!+\!
\left (\!i{q \over m_e}\!-\!\omega_c\!\times\! {\p v^\Sigma \over m_e}\right )\!
\delta \Sigma_{i}\p vg_{i}
\nonumber\\
&S_i={iq\p v g_{i}\over m_e} \Phi+2(\delta \Sigma\times g)_i+{\delta n \partial_\mu g\over \tau\partial_\mu n}e_i
\nonumber\\
&
\!+\!\left (\!i{q \over m_e}\!+\!\omega_c\!\times\! {\p v^\Sigma \over m_e}\right )\!\delta \Sigma_0\p v g_{i}\!+\!
\left (\!i{q \over m_e}\!-\!\omega_c\!\times\! {\p v^\Sigma \over m_e}\right )\!\delta \Sigma_{i}\p v f.
\label{source1}
\end{align}

Compared with the result without magnetic field we see that the source terms 
(\ref{source}) get additional rotation terms coupled to the momentum-dependent 
derivation of meanfields (\ref{deltaS}) which is present only with extrinsic spin-orbit coupling. Further the drift side gets an explicit derivative with respect to the velocity which we will take into account in the following.

\subsection{Solution of linearized equations}
In order to solve (\ref{lina}) and (\ref{linb}) we use the same coordinate
system as Bernstein \cite{Be58}. The magnetic field $\V B$ points in the $v_z$ direction and the
$q$ vector is in the $v_z-v_x$ plane with an angle $\Theta$ between $v_x$ and
$q$
\be
\V q&=&q\sin\Theta \V e_x+q\cos \Theta \V e_z.
\label{def1a}
\ee
For the velocity $v$ we use polar coordinates around $\V B$ with an azimuthal angle $\phi$
\be
\V v(\phi)&=&w\cos\phi \V e_x+w \sin\phi \V e_y+u\V e_z
\label{def1}
\ee
and one gets
\ba
&{1\over \tau}\!-\!i\omega\!+\!i\V q\V v\!+\!\left (\V v\times {e \V B\over m}\right )\V  {\p v}={1\over \tau}\!-\!i\omega\!+\!i\V q\cdot
\V v(\phi)\!-\! \omega_c\p \phi\nonumber\\
&
={1\over \tau}-i\omega+i\V q\cdot
\V v(\omega_c t_\phi) -\p t\equiv -i\Omega_{t_\phi}-\p {t_\phi}
\label{def2}
\end{align}
with the orbiting time 
\be
t_\phi=\phi/\omega_c.
\label{corresp}
\ee 
We can write the equations (\ref{lina}) and (\ref{linb}) as
\begin{align}
&\left (-i\Omega_{t_\phi}-\p {t_\phi}\right ) \delta f+\left ({iq\p v\V \Sigma\over m_e}+\V \tau^{-1}\right )\cdot \V \delta g=S_0\nonumber\\
&\left (\!-\!i\Omega_{t_\phi}\!-\!\p {t_\phi}\right ) \delta g_{i}\!+\!\left ({iq\p v\Sigma_{i}\over m_e}\!+\!(\V \tau^{-1})_i\right ) \delta f\!-\!2 (\V \Sigma\times \V \delta g)_i=S_{i}.
\label{lin}
\end{align}
where the corresponding right hand sides are given by (\ref{source1}). Now we
employ the identity
\be
&&\V {\cal B}\cdot \delta \V g+ (\V \sigma \cdot \V {\cal B}) \delta f-2 \V \sigma \cdot (\V \Sigma\times \delta \V g)=
\nonumber\\&&
\V \sigma \cdot {\V {\cal B}+2 i\V \Sigma\over 2} \delta \hat F+\delta \hat F\V \sigma \cdot {\V {\cal B}-2 i\V
  \Sigma\over 2}
\ee
with $\V {\cal B}=i q\p v \V\Sigma/m_e +\V \tau^{-1}$ and $\delta \hat F=\delta f+\V \sigma\cdot \delta \V g$ which one proves with the help of $(\tau a)(\tau b)=a\cdot b+i \tau (a\times b)
$. 
This allows to rewrite (\ref{lin}) into 
\ba
&-\p {t_\phi} \hat \delta F-i\Omega_{t_\phi} \hat \delta F\nonumber\\
&+\V\sigma \cdot  \left ({\V {\cal B}\over 2} +i \V \Sigma\right )\delta \hat F+\delta \hat F\,\V \sigma \cdot \left ({\V {\cal B}\over 2} -i \V \Sigma\right )=\hat S_{p_\phi}(\omega)
\label{linc}
\end{align}
where $\hat S_{p_\phi}=S_0+\V \sigma\cdot \V S$.

Please note that due to (\ref{corresp}) the integration over the 
azimuthal angle is translated into the time integration about orbiting intervals.
Therefore Eq. (\ref{linc}) has a great similarity to the time-dependent Eq. (\ref{opS}).

Equation (\ref{linc}) is easily solved as
\ba
&\delta F=
-\!\int\limits_{\infty}^t \! d\bar t {\rm e}^{i\int\limits_t^{\bar
    t}\Omega_{t'}^+
d t'}{\rm e}^{\V \sigma
  \int\limits^t_{\bar t}\left ({\V {\cal B}_{t'}\over 2} \!+\!i \V \Sigma_{t'}\right ) d t'}
\!\hat S_{p_{\omega_c \bar t}}
{\rm e}^{\V \sigma
  \int\limits^t_{\bar t}\left ({\V {\cal B}_{t'}\over 2} \!-\!i \V \Sigma_{t'}\right )d t'}\nonumber\\
&=\int\limits_{-\infty}^0 d x 
{\rm e}^{-i\int\limits_0^{x}\Omega_{t\!-\!y}^+d y}
\nonumber\\
&\times
{\rm e}^{\V \sigma \int\limits_0^{x}\left (\frac 1 2{\V {\cal B}_{t-y}} +i \V \Sigma_{t-y}\right )d y}
\hat S_{p_{\omega_c (t\!-\!x)}}(\omega)
{\rm e}^{\V \sigma  \int\limits_0^{x}\left (\frac 1 2{\V {\cal B}_{t-y}} -i \V \Sigma_{t-y}\right )d y}
\label{sol}
\end{align}
where we used $\omega^+=\omega+i \tau^{-1}$ as before.
The first exponent can be calculated explicitly with the definitions of
(\ref{def1}) and (\ref{def2})
\be
i\int\limits_0^{x}\Omega_{t-y}^+d y=i \omega^+x- i \V q \cdot {\cal R}_{x} \cdot \V v(t)
\ee
with the matrix \cite{WZT05}
\be
{\cal R}_x=\frac{1}{\omega_c}\left (\begin{matrix} 
{\sin\omega_cx}&{1-\cos \omega_c x}&0\cr
{\cos\omega_cx-1}&{\sin \omega_c x}&0\cr
0&0&\omega_c x
\end{matrix}
\right )
\ee
having the property ${\cal R}_{-x}=-{\cal R}_x^T$.
Neglecting the magnetic-field dependence in the phase $qRv=q v+o(B)$ we obtain with (\ref{sol}) exactly again the solution (\ref{251}) but with an additional retardation in the momentum $S_{p_{\omega_c (t\!-\!x)}}$ instead of $S_{p_{\omega_c t}}=S_p$ in (\ref{251}).

%
%
%
We employ the long-wavelength approximation $\V {\cal B}\approx 0$ neglecting
the vector relaxation. The integration over an azimuthal angle $x=\phi/\omega_c$ is coupled to the momentum (velocity) arguments. The spin-orbit coupling provides a momentum-dependent $\V \Sigma$ which couples basically to $q\p p \V \Sigma$. Since
\ba
m_e q\p p=q\sin\Theta\left (\cos\phi\p w-{\sin\phi\over w} \p \phi\right )+q \cos\Theta \p u
\end{align}
in the coordinates (\ref{def1a}) and (\ref{def1}), it means we neglect higher than first order derivatives in $\phi$ and $\partial_p\V \Sigma$ when approximating $\V \Sigma_{t-y}\approx \V \Sigma_t$ in the exponent.
We obtain
\ba
\delta \hat F=\delta f\!+\!\V \sigma\cdot \delta \V g=\!\!\!\int\limits_{-\infty}^0 \!\!\!d x
{\rm e}^{i (\V q {\cal R}_x\V v_t\!-\!\omega^+ x)}
{\rm e}^{ix \V \sigma\V \Sigma_{t}}\hat S_{p_{\omega_c (t\!-\!x)}}{\rm e}^{-i x \V \sigma \V \Sigma_t}.
\label{solutf}
\end{align}
To work it out further we  use
$
{\rm e}^{i\tau\cdot a}=\cos|a|+i{\tau\cdot a\over |a|}\sin|a|
$
to see that
\ba
&{\rm e}^{i\V \sigma \cdot \V \Sigma x}(S_0+\V \sigma \cdot \V S){\rm e}^{-i \V \sigma \cdot \V \Sigma x}=
S_0+(\V \sigma\cdot \V S) \cos(2 x|\Sigma|)\nonumber\\&
+\V \sigma (\V S\times \V e)\sin(2 x |\Sigma|)+{(\V \sigma \cdot \V e)(\V S\cdot \V e)}(1-\cos(2 x |\Sigma|))
\label{3}
\end{align}
with the direction $\V e=\V \Sigma/|\Sigma|$ and (\ref{sig}).

The effect of a magnetic field is basically condensed at two places. First the phase term $\V q \cdot {\cal R}_{x} \cdot \V p= \V q\cdot \V p+o(B)$  and we have $\bar \omega=\omega -\V p\cdot \V q/m+i/\tau$
\ba
&\int\limits_{-\infty}^0 {\rm e}^{-i(\omega x\!-\!\V q \cdot {\cal R}_{x} \cdot {\V p\over m})} \begin{pmatrix} \!\cos 2|\Sigma| x\cr 1\!-\!\cos 2|\Sigma| x \cr -\sin 2|\Sigma| x\!\end{pmatrix}
\nonumber\\&
=
\frac 1 2 \!\begin{pmatrix}\!{i\over \bar \omega \!+\!2 \Sigma}\!+\!{i\over \bar \omega\!-\!2 \Sigma}\cr
{2 i\over \bar \omega}\!-\!{i\over \bar \omega \!+\!2 \Sigma}\!-\!{i\over \bar \omega\!-\!2 \Sigma}\cr
{1\over \bar \omega \!-\!2 \Sigma}\!-\!{1\over \bar \omega\!+\!2 \Sigma}\!\end{pmatrix}\!+\!o(q^2,B)
=
\begin{pmatrix}
{i\over \bar \omega}\cr 0\cr {2 \Sigma \over \omega^2}
\end{pmatrix}\!+\!o(B,\Sigma^2).
\label{costime1}
\end{align}
The magnetic-field-dependent phase factor ${\cal R}_x$ does play a role only in inhomogeneous systems with finite wavelength. In the limit of large wavelength this effect can be ignored.

The $sin$ and $cos$ terms are results of the precession of spins around the effective direction $\V e=\V \Sigma/\Sigma$ and can be considered as Rabi oscillations. For the limit of small $\Sigma$ we can expand the $cos$ and $sin$ terms in first order
$\approx S_0+\V \sigma \cdot \V S-2\V \sigma \cdot (\V S\times \V \Sigma)x$ as was analyzed in \cite{M14}.
The second effect is the retardation in $t=\phi/\omega_c$ which means that the precession time in the arguments $S(t-x)$ contains important magnetic field effects. In fact, this retardation represents all kinds of normal Hall effects as we will convince ourselves now.

\subsection{Retardation subtleties by magnetic field}

The magnetic field causes a retarding integral in the last section over the precession time $t=\phi/\omega_c$ coupled to any momentum by the representation in Bernstein coordinates
\be
{\V p_\phi\over m}=(w\cos\phi, w\sin \phi, u) .
\ee
This retardation is crucial for any kind of Hall effect. In order to get a
handle on such expressions we concentrate first on the mean values of the scalar part $\delta f$. The general field-dependent solution provides a form
\be
\langle A \rangle=\sum\limits_{p_\phi}\int\limits_{-\infty}^0dx {\rm e}^{-i(\omega^+ x-\V q {\cal R}_x {\V p_\phi\over m})} A(\V p_{\phi})S_0(p_{\phi-x \omega_c})
\ee
where $S_0$ is the scalar source term.
The trick is to perform first a shift $\phi\to\phi+\omega_c x$ and integrate then about
$p=p_\phi$. This has the effect that the retardation is only condensed in the momentum of variable $A$
\ba
&{\cal P}(x)=\V p_{\phi+\omega_c x}
\nonumber\\&
=\V p_\phi \cos(\omega_c x)\!+\!\V e_z(\V e_z\cdot \V p_\phi) [1\!-\!\cos(\omega_c x)]
\!+\!\V e_z\!\times\!\V p_\phi \sin(\omega_c x)
\nonumber\\&
=
\V p_\phi+\V e_z\times\V p_\phi \omega_c x+o(\omega_c^2).
\label{pphi}
\end{align}
and the exponent
\be
{\cal R}_x {\V p_{\phi+\omega_c x}\over m}&=&{\V p_\phi \over m} {\sin\omega_c x\over \omega_c}+\left (\V e_z\times {\V p_\phi\over m}\right ){1-\cos\omega_c x\over \omega_c}
\nonumber\\
&=& {\V p_\phi\over m} x+\omega_c {x^2\over 2} \V e_z\times {\V p_\phi\over m}+o(\omega_c^2).
\ee
The phase effect leads to the first order corrections in $\omega_c$ or alternatively in wave length $q$ 
\ba
&\langle A \rangle=\sum\limits_{p} S_0(p)\,
\left [1-{\omega_c \over 2 m}\V q\cdot (\V e_z\times \V p) \, \partial^2_\omega +o(\omega_c^2)\right ]
\nonumber\\&\times 
\int\limits_{-\infty}^0dx {\rm e}^{-i(\omega^+ x-{\V q\cdot \V p\over m})}
A\left [{\cal P}(x)\right ]
\nonumber\\
&=\sum\limits_{p} S_0(p)\,
A\left [{\cal P}(i\p \omega)\right ]  
\nonumber\\&\times 
\left [1-{\omega_c \over 2 m}\V q\cdot (\V e_z\times \V p) \, \partial^2_\omega +o(\omega_c^2)\right ]
{i\over \omega^+ -{\V q \cdot \V p\over m}}.
\label{mean0}
\end{align}
where the integration variable $x$ in the momentum (\ref{pphi}) can be transformed into derivatives of $\omega$ if needed.

Completely analogously we can perform any mean value over the vector part of the distribution $\V \delta g$. We have
\ba
&\langle \V A \rangle=\sum\limits_p A(p)\delta \V g=
\sum\limits_{p}\left [1-{\omega_c \over 2 m}\V q\cdot (\V e_z\times \V p) \, \partial^2_\omega +o(\omega_c^2)\right ]
\nonumber\\
&\times\int\limits_{-\infty}^0\!\!\!dx {\rm e}^{i({\V q \V p\over m}x\!-\!\omega^+ x\!+\!o(\omega_c, q^2))}A[{\cal P}(x)] 
\nonumber\\
&\times \!\left [\! \V S \cos (2 \Sigma x)\!+\!\V e\!\times\! \V S \sin(2 \Sigma x)\!+\!\V e (\V e \cdot \V S)(1\!-\!\cos(2\Sigma x))\right ]
\end{align}
where the arguments of $\V S$, $\Sigma$ and $\V e$ are the momentum $p$ and no retardation anymore. The exponent can be written in complete B-dependence with ${\cal R}_x$ of course. Then the x-integration over the $cos$ and $sin$ terms has to be performed numerically. Analytically we can proceed if we expand the phase effect in orders of $\V q$.
We obtain with the help of (\ref{costime1}) 
\ba
&\langle \V A \rangle=
\sum\limits_{p}A[{\cal P}(i\p \omega)]\left [1-{\omega_c \over 2 m}\V q\cdot (\V e_z\times \V p) \, \partial^2_\omega +o(\omega_c^2)\right ]
\nonumber\\
&\times
\left [ 
[\V e \times (\V S\times \V e)] \frac i 2 \left ({1\over \bar \omega +2\Sigma}+{1\over \bar \omega -2\Sigma}\right)
\right .\nonumber\\
&\left .+\V e\times \V S \frac 1 2 \left ({1\over \bar \omega + 2 \Sigma}-{1\over \bar \omega -2\Sigma}\right )
+\V e(\V e\cdot \V S) {i\over \bar \omega}
\right ]
\label{ruleM}
\end{align}
with $\bar \omega=\omega+{i\over \tau}-{\V p\V q\over m}$ and (\ref{pphi}).

The formula (\ref{mean0}) and (\ref{ruleM}) establish the rules for calculating mean values with magnetic fields.
The usefulness of these rules can be demonstrated since it simplifies the way to obtain the linearized solutions (\ref{ds1}), (\ref{ds2}) and (\ref{ds3}) tremendously. In fact integrating with $A=1$ we obtain straightforwardly the response functions and the equation system (\ref{final}). This shows that up to linear order in wave-vector the magnetic field enters only via the Zeeman term in $\V \Sigma$.

\subsection{Classical Hall effect}

Now, we are in a position to see how the Hall effect 
is buried in the theory. Therefore we neglect any meanfield and spin-orbit
coupling for the moment such that the f and g distributions decouple and use
the $q\to 0$ limit, i.e. homogeneous situation. We obtain from (\ref{sol}) with (\ref{3}) and (\ref{source1})
\be
\delta f=  -\int\limits_{-\infty}^0 d x {\rm e}^{-i \omega^+ x
} {e \V E}\cdot \V {\p {p_\phi}} f(p_{\phi-\omega_c x})
\ee
where we now pay special care to the retardation since this provides the Hall effect which was overseen in many treatments of magnetized plasmas.

After the shift of coordinates in azimuthal angle $\phi$ as outlined in the last section, we can carry out the $x$-integration with the help of (\ref{pphi}), (\ref{mean0}) and (\ref{costime1}): 
\be
\V J &=&e \sum\limits_{p_\phi} {\V {p_\phi}\over m_e} \, \delta f
\nonumber\\
&=& 
-e^2\sum\limits_{p_\phi}\V E\cdot \V {\p {p_\phi}} f(p_\phi) \int\limits_{-\infty}^0 d x {\rm e}^{-i \omega^+ x} {\V p_{\phi+\omega_c x}\over m}
\nonumber\\
&=&
\sigma_0 {1-i\omega \tau\over (1-i\omega \tau)^2+(\omega_c \tau)^2 }
\left [\V E+{(\omega_c \tau)^2\over (1 -i \omega \tau)^2} (\V E\cdot \V e_z)\V e_z
\right .\nonumber\\
&&\left . +{\omega_c \tau\over 1 -i \omega \tau} \V E\times \V e_z \right ]
\label{Hallclass}
\ee
which agrees of course with the elementary solution of
\be
m_e \dot {\V v}=e (\V v\times \V B)+e \V E-m_e {\V v\over \tau}.
\label{start}
\ee

In order to obtain all three precession terms we have used the complete form (\ref{pphi}) and no expansion in $\omega_c$.

\subsection{Quantum Hall effect}

If we consider low temperatures such that the motion of electrons become
quantized in Landau levels we have to use the quantum kinetic equation and not
the quasi-classical one. However, we can establish a simple {\it
  re-quantization rule} which allows us to translate the above discussed
quasi-classical results into the quantum expressions. Therefore we recall the
linearization of the quantum-Vlasov equation, which is the quantum kinetic
equation with only the mea-field in operator form:
\be
\dot \rho-\frac i \hbar [\rho, H]=0.
\ee
The perturbing Hamiltonian due to external electric fields is $\delta H=e \V E\cdot  \V {\hat x}$ such that the linearization in eigenstates $E_n$ of the unperturbed Hamiltonian reads
\be
\delta \rho_{nn'}=-e\V E \cdot \V x_{nn'} {\rho_n-\rho_{n'}\over \hbar \omega -E_n+E_{n'}}.
\ee  
One obtains the same result in the vector gauge since $[\rho,\delta H]={e \V E t\over m_e}[\rho, \V p]=e  \V E[\rho,\V v]$ and the same matrix elements appear.
Now, we investigate the quasi-classical limit where the momentum states are proper representations. We chose
\be
\bra n=\bra {p_1}=\bra {p+\frac q 2},\quad \ket n'=\ket {p_2}=\ket {-p+\frac q 2}
\ee
and we have in the quasi-classical $q\to 0$ approximation
\ba
\V x_{nn'} (\rho_n-\rho_{n'})&=\frac \hbar i \V {\p q}\delta (\V q) (\rho_{p+\frac q 2}-\rho_{p-\frac q 2})
\nonumber\\
&\approx  \frac \hbar i \V {\p q}\delta (\V q)\V q \cdot \V {\p p} \rho_p
=-\frac \hbar i \delta (\V q) \V {\p p} \rho_p
\end{align}
from which follows
\be
\delta \rho\approx -i\hbar {e \V E \cdot \V{\p p} \rho\over \hbar \omega -{\V p\cdot \V q\over m_e}}. 
\ee
This is precisely the quasi-classical result we obtain from quasi-classical kinetic equations. Turning the argument around we see that we can re-quantize our quasi-classical results by applying the rule
\be
\V E\cdot \V {\p p} f\to \V E\cdot \V v_{nn'}{f_n-f_{n'}\over E_{n'}-E_n}.
\ee

Let us apply it to the normal Hall conductivity. We use the area density $1/A$ and re-normalize the level distribution $\sum_n f_n=1$ to obtain for the static conductivity $\omega=0$
\be
\sigma_{\alpha\beta}={e^2\hbar i\over A}\sum\limits_{nn'} f_n (1\!-\!f_{n'}) {1\!-\!{\rm e}^{\beta (E_n\!-\!E_{n'})}\over (E_n\!-\!E_{n'})^2} v_{nn'}^\alpha v_{n'n}^\beta
\label{si1}
\ee
which is nothing but the Kubo formula
. Further evaluation for Landau levels has been performed by Vasilopoulos \cite{V85,VV88}. Therefore one chose the gauge $\V A=(0, B x,0)$ and the corresponding energy levels are
\be
E_n=\left (n+\frac 1 2\right )\hbar \omega_c +{p_z^2\over 2 m_e}
\ee
where the last term is only in 3D. The wave functions read
\be
\ket n={1\over \sqrt{A}}\phi_n(x+x_0) {\rm e}^{i p_y y/\hbar}{\rm e}^{i p_z z/\hbar}
\ee 
with the harmonic oscillator functions $\phi_n$, $x_0=l^2 p_y/\hbar$, $l^2=\hbar/e B$,  and $A=L_y L_z$ where the corresponding $z$ parts are absent in 2D. The calculation in 3D can be found in \cite{VV88}. Here, we represent the 2D calculation. One easily obtains
\be
\!\!v_{nn'}v_{n'n}\!=\!{i\hbar \omega_c\over 2 m}\!\left [n \delta_{n',n\!-\!1}\!-\!(n\!+\!1)\delta_{n',n\!+\!1}\right ]\delta(p_y\!-\!p_y').
\ee
Introducing this into (\ref{si1}) and using
\be
\sum\limits_{p_y}={L_y\over 2  \pi \hbar} \int\limits_{-{\hbar L_x\over 2 l^2}}^{{\hbar L_x\over 2 l^2}} d p_y ={A\over 2 \pi l^2}
\ee
one arrives at
\ba
{e^2\over h}\sum\limits_{n'}(n'+1)f_{n'}(1-f_{n'+1})\left (1-{\rm e}^{-\beta \hbar \omega_c}\right )\to {e^2\over h}(\bar n+1) 
\end{align}
with $\bar n \omega_c\le \epsilon_f\le (\bar n+1) \omega_c$. This is von Klitzing's result for $T\to 0$.

\subsection{Polarization functions}

Integrating (\ref{sol}) over the momentum $p=v m_e$ and solving algebraically for $\delta n$ and $\delta s$
one gets the response functions (\ref{final}) with the B-field
modifications. This concerns the precession-time integration instead of the
energy denominator coupled to the tensor $q{\cal R}p$ and retardations in the
momentum integration as described above. Especially with the help of
(\ref{ruleM}) the discussed polarizations (\ref{abbrev})-(\ref{pol1}) can be
easily translated with (\ref{costime1}) such that the effect of the magnetic
field in the phase can be considered. The retardations does not play any role
since for the density and spin response we do not have moments of momentum
that would be retarded. The numerical results of these phase effects for small
$\Sigma$ have been discussed in \cite{M14} leading to a staircase structure of
the response functions with respect to the frequency at Landau levels. The excitation shows a splitting of the collective mode into Bernstein modes. Since it was presented in \cite{M14} the repetition of results is avoided here.

\section{Summary}

We have solved the linearized coupled kinetic equations for the density and
spin Wigner distributions in arbitrary magnetic fields, vector and scalar
meanfields, spin-orbit coupling and relaxation time approximations obeying the
conservation of density. The response functions for the density and spin
polarization with respect to an external electric field are derived. Various
forms of polarization functions appear reflecting the complicated nature of
different precessions and including Rabi shifts due to the effective Zeeman
field. The latter consists of the magnetic field, the magnetization due to
impurities, the spin polarization, and the spin-orbit coupling. 

The long wavelength expansions are presented and the density and spin
collective modes are determined for neutral and charged scattering
separately. For a neutral system,  no optical charge mode appears but there
are three optical spin modes. These are dependent on the spin-orbit coupling and the effective Zeeman field. The energy and damping of these modes are found to be linearly dependent on the spin-orbit coupling.
A spin-wave instability is reported and the range where such spin segregation can appear are calculated. 

The charge and spin waves for charged Coulomb scattering show that only
transverse spin modes can exist with respect to an effective magnetization
axis. The charge density waves are damped plasma oscillations and the spin
waves are splitting into two modes dependent on the polarization. One mode
decreases in energy and becomes damped with increasing polarization while the
second mode increases and becomes sharper again with increasing
polarization. This analysis was possible with the help of the polarization,
magnetic field and spin-orbit dependent dielectric function which was
presented here as a new result. The range of instability with respect to
frequency, polarization and collisional damping is presented which is again
interpreted as spin segregation. The latter view is supported by the
discussion of the statical and dynamical screening length  whose dependence on
the polarization and spin-orbit coupling is derived.

Finally, the spin response shows an interesting damped oscillation behavior
different in each direction originating from the off-diagonal responses. The
magnetic field causes an ellipsoidal relaxation which shows a rotation of the
polarization axes depending on the spin-orbit coupling. We find a cross-over
from damped oscillation to exponentially decay dependent on the polarization
and collision frequency. Spin segregation as a consequence is discussed and the dephasing times are extracted.

The response with an external magnetic field shows some subtleties in
retardations when observables of the Wigner functions are calculated. In fact,
the Hall effect is possible to obtain only when these retardations are taken
into account. The quantum version of the quasi-classical kinetic equations is
shown to provide the quantum-Hall effect. Explicit calculations of the
response function show a staircase behavior with respect to the frequency at the Landau levels. At these frequencies the Rabi satellite response functions become large leading to out-of-plane resonances \cite{M14}.

\appendix

\section{Solving vector equations}

In the following all symbols are vectors and we search for solutions $y$ in
terms of capital symbols. 
We start with the simplest vector equation
\be
y_1=B-Q(V \cdot y_1)
\label{1e}
\ee
which is easily solved by iteration and  the geometrical sum
\be
y_1=B-Q (V\cdot B){1\over 1+Q\cdot V}={B+V\times (B\times Q)\over 1+ Q\cdot V}.
\label{1s}
\ee

Next we consider the equation of the type
\be
y_2=B-A\times y_2
\label{2e}
\ee
which by iterating once leads to
\be
(1+A^2)y_2=B-A\times B+A(A\cdot y_2)
\ee
which is again of the type (\ref{1e}) with $B\to (B-A\times B)/(1+A^2)$, $V\to-A$, and
$Q\to A/(1+A^2)$ such that the solution reads
\be
y_2={B-A\times B+A(A\cdot B)\over 1+A^2}.
\label{2s}
\ee
The combined type reads
\be
y_3+A\times y_3=B-Q(V \cdot y_3)
\label{3e}
\ee
where in a first step we consider the right hand side as a $B$ of the problem
(\ref{2e}) and get the
solution according to (\ref{2s}). This leads to the problem (\ref{1e}) with
$B\to (B-A\times B+A (A\cdot B))/(1+A^2)$ and $Q\to (Q-A\times Q+A(A\cdot
Q))/(1+A^2)$ such that the solution can be written according to (\ref{1s})
\be
y_3&=&{B-A\times B+A(A\cdot B)+V\times H\over 1+A^2+Q\cdot V-V(A\times
  Q)+(A\cdot V)(A\cdot Q)}\nonumber\\
H&=&(B-A\times B+A(A\cdot
  B))\times{Q-A\times Q+A(A\cdot Q)\over 1+A^2}
\nonumber\\
&=&B\times Q+A\times(Q\times B)
\ee
where the last equality is a matter of algebra. The final solution
reads therefore
\ba
&y_3=\nonumber\\
&{B-A\times B+A(A\cdot B)+V\times [B\times Q+A\times(Q\times B)]\over
  1+A^2+Q\cdot V-V\cdot (A\times
  Q)+(A\cdot V)(A\cdot Q)}\nonumber\\
&\equiv{y_{3z}\over y_{3n}}.
\label{3s}
\end{align}

As a next complication we consider the vector equation where the scalar products
appear with respect to two vectors
\be
y_4+A\times y_4+Q(V \cdot y_4)=B-P(T\cdot y_4)
\label{4e}
\ee
which is recast into the problem (\ref{3e}) with $B\to B-P(T\cdot y_4)$ such
that we obtain
\ba
&y_4=y_3-(T\cdot y_4)Q_1\nonumber\\
&Q_1=\nonumber\\
&{P+P\times A+V\times (A\times (Q\times P)-Q\times
  P)+A(A\cdot P)\over y_{3n}}
\end{align}
which is the problem (\ref{1e}) with $V\to T$, $B\to y_3$ and $Q\to Q_1$
such that we obtain
\be
y_4={y_3-T\times (Q_1\times y_3)\over 1+T\cdot Q_1}.
\ee
The cross product in the numerator can be shown by somewhat lengthy
calculation to be
\be
&&Q_1\times y_3={P\times B+A\times (B\times P)+V (B\cdot P\times Q)\over y_{3n}}\nonumber\\&&
\ee
such that the final solution reads
\ba
&y_4=\nonumber\\
&\frac{y_{3z}-T\times [A\times (B\times P)-B\times P+V P\cdot (B\times Q)]}
{y_{3n}+T\cdot y_{4h}}\nonumber\\
&
y_{4h}=\nonumber\\
&P\!+\!P\times A\!-\!V\!\times\! [A\!\times\! (P\!\times\! Q)+P\!\times
\!  Q]\!+\!A (A\cdot P)
\label{4s}
\end{align}

\section{Evaluation of operator forms}\label{operator}

In this appendix we evaluate the operator form
\ba
&\int\limits_0^\infty d x\,  {\rm e}^{i\bar \omega x}
{\rm e}^{\left({\V b}+\V a\right)\cdot \V \sigma x} 
(S_0+\V \sigma \cdot \V S) 
{\rm e}^{\left({\V b}- \V a\right)\cdot \V \sigma x}
\nonumber\\&
\label{c1}
\end{align}
with $\bar \omega=\omega+i/\tau$.
First we rewrite the exponential of Pauli matrices to obtain
\be
{\rm e}^{\left({\V b}\pm\V a\right)\cdot \V \sigma x}=\cos c_\pm x+i \V e_\pm \cdot \V \sigma \sin c_\pm x
\label{c2}
\ee
where we use
\be
c_\pm=|b\pm a|;\qquad \V e_\pm={\V b\pm \V a\over |\V b\pm \V a|}.
\ee
To evaluate the occurring products it is useful to deduce from $(\V a\cdot \V \sigma)(\V b\cdot \V \sigma)=\V a\cdot \V b+i \V \sigma \cdot (\V a\times \V b)$ the relations
\be
\V \sigma \cdot (\V a \cdot \V \sigma)=\V a+i (\V a\times \V \sigma),\quad 
(\V a \cdot \V \sigma)\cdot \V \sigma  =\V a-i (\V a\times \V \sigma)
\ee
with the help of which we find
\ba
&(\sigma \cdot \V a)(\V b \times \V \sigma)
=
-((\sigma \cdot \V a)\V \sigma\times \V b)
=-(\V a-i (\V a\times \V \sigma))\times \V b
\nonumber\\
&=-\V a\times \V b+i \V b \times (\V \sigma\times \V a)
=-\V a \times \V b+i \V \sigma (\V a \cdot \V b)-i \V a (\V b\cdot \V \sigma).
\end{align}
One obtains from (\ref{c2})
\ba
&{\rm e}^{\left({\V b}+\V a\right)\cdot \V \sigma x} 
{\rm e}^{\left({\V b}- \V a\right)\cdot \V \sigma x}=
\cos c_+x\cos c_-x
\nonumber\\
&
+i (\V e_+\cdot \V \sigma) \sin c_+x \cos c_-x
+i (\V e_-\cdot \V \sigma) \sin c_-x \cos c_+x
\nonumber\\
&
-[\V e_+\cdot \V e_-+i \V \sigma \cdot (\V e_+\times \V e_-)] \sin c_+x\sin c_-x
\label{c6}
\end{align}
and
\ba
&{\rm e}^{\left({\V b}+\V a\right)\cdot \V \sigma x} 
\V \sigma {\rm e}^{\left({\V b}- \V a\right)\cdot \V \sigma x}=
\V \sigma \cos c_+x\cos c_-x
\nonumber\\
&
\!+\!i (\V e_+\!-\! \V \sigma\times \V e_+) \sin c_+x \cos c_-x
\nonumber\\
&\!+\!i (\V e_-\!+\! \V \sigma\times \V e_+) \sin c_-x \cos c_+x
\nonumber\\
&
\!+\![\V \sigma (\V e_+\cdot \V e_-)\!-\! (\V \sigma \cdot \V e_+)\V e_- 
\!-\! (\V \sigma \cdot \V e_-)\V e_+\!+\!i (\V e_+\times \V e_-)]
\nonumber\\
&\quad\times \sin c_+x\sin c_-x.
\label{c7}
\end{align}
Now we evaluate the integrals over the cos and sin functions.
Due to the positive imaginary part of $\omega$ the integral vanishes at the upper infinite limit and one has
\be
\int\limits_0^\infty d x\,  {\rm e}^{i\omega x}\cos (c x)&=&{i\omega \over \omega^2-c^2}
\nonumber\\
\int\limits_0^\infty d x\,  {\rm e}^{i\omega x}\sin (c x)&=&-{c \over \omega^2-c^2}.
\ee
We will need 
\be
(c_+\pm c_-)^2=2 (a^2+b^2\pm |a^2-b^2|).
\ee
which leads to either to $4 a^2$ or $4 b^2$ dependent whether $a^2\gtrless b^2$ and the $\pm$ sign respectively. Therefore one obtains 
\ba
&\int\limits_0^\infty d x\,  {\rm e}^{i\omega x}\cos (c_+ x)\cos (c_- x)
={i\omega \over 2} 
\left ({1\over \omega^2\!-\!4 b^2}\!+\!{1\over \omega^2\!-\!4 a^2} \right )
\nonumber\\
&\int\limits_0^\infty d x\,  {\rm e}^{i\omega x}{\sin (c_\pm x)\sin (c_\mp x)\over |a^2-b^2|}
=-{2 i\omega  \over (\omega^2\!-\!4 a^2)( \omega^2\!-\!4 b^2)}
\nonumber\\
&\int\limits_0^\infty d x\,  {\rm e}^{i\omega x}
\left (
{\sin (c_+ x)\cos (c_- x)\over |\V a+\V b|}
+
{\sin (c_- x)\cos (c_+ x)\over |\V a-\V b|}
\right )
\nonumber\\
&\qquad
=-{2 \omega^2  \over (\omega^2\!-\!4 a^2)( \omega^2\!-\!4 b^2)}
\nonumber\\
&\int\limits_0^\infty d x\,  {\rm e}^{i\omega x}
\left (
{\sin (c_+ x)\cos (c_- x)\over |\V a+\V b|}
-
{\sin (c_- x)\cos (c_+ x)\over |\V a-\V b|}
\right )
\nonumber\\
&\qquad
={8 \V a\cdot \V b  \over (\omega^2\!-\!4 a^2)( \omega^2\!-\!4 b^2)}
.
\end{align}

This allows to calculate the different occurring integrals in (\ref{c1}) with 
(\ref{c6}) and (\ref{c7}) as
\ba
&\int\limits_0^\infty d x\,  {\rm e}^{i\omega x}
{\rm e}^{\left({\V b}+\V a\right)\cdot \V \sigma x} 
{\rm e}^{\left({\V b}- \V a\right)\cdot \V \sigma x}
\nonumber\\
&={4\omega \V \sigma \cdot (\V b\!\times\! \V a)\!+\!i \omega (\omega^2\!-\!4 a^2)\!+\!8 i (\V \sigma \cdot \V a)(\V a \cdot \V b)\!-\!2 i \omega^2 (\V \sigma \cdot \V b)\over (\omega^2\!-\!4 a^2)( \omega^2\!-\!4 b^2)}
\label{c11}
\end{align}
and
\ba
&\int\limits_0^\infty d x\,  {\rm e}^{i\omega x}
{\rm e}^{\left({\V b}+\V a\right)\cdot \V \sigma x} 
\V \sigma 
{\rm e}^{\left({\V b}- \V a\right)\cdot \V \sigma x}
=\biggl \{
4\omega (\V a\!\times\! \V b)\!+\!8 i \V a (\V a \cdot \V b)
\nonumber\\
&
\!+\!
i \omega [
\V \sigma (\omega^2\!-\!4 b^2)
\!+\!4 \V b(\V \sigma \cdot \V b)\!-\!4 \V a (\V \sigma \cdot  \V a )
\!-\!2 \V b \omega]
\nonumber\\&
\!+\!8 (\V a \cdot \V b)(\V b \times \V \sigma)\!+\!2 \omega^2(\V \sigma \times \V a)
\biggr \}
{1 \over (\omega^2\!-\!4 a^2)( \omega^2\!-\!4 b^2)}.
\label{c12}
\end{align}

\section{Long wavelength expansion\label{expansion}}  

In order to discuss dispersion relations and collective modes we need the expansion of all polarization functions up to second order in wavelength appearing in terms of
\ba
\Pi\left (\omega+{i\over \tau}-{\V p\cdot \V q \over m_e}\right )=
\left (1
-{\V p\cdot \V q\over m_e} \p \omega 
+{(\V p\cdot \V q)^2\over m_e^2} \partial_\omega^2 \right ) \Pi(\omega_+)
\label{ex1}
\end{align}
where we use $\omega_+=\omega+i/\tau$. A further wave-length term comes from the magnetic field dependence of the polarization function discussed in chapter IV.C which leads to a term linear in the magnetic field
\be
1-{\omega_c\over 2 m_e} \V q\cdot (\V e_z\times \V p )\partial_\omega^2.
\label{ex2}
\ee

Any function of $\Sigma$ we can expand therefore as
\be
\Pi(\Sigma)\!=\!\left [1\!+\!\left ({b_\perp^2\over 2}\!+\!b_3\right )\Sigma_n\partial_{\Sigma_n}\!+\!{b_3^2\over 2}\Sigma_n^2\partial_{\Sigma_n}^2\right ] \Pi(\Sigma_n).
\label{ex3}
\ee

Summarizing we have to apply (\ref{ex1}), (\ref{ex2}) and (\ref{ex3}) to all
polarization function and calculate the momentum integration. Here we give the
final results which may be obtained after some lengthy calculation. Since
$b_\perp(\V p)$ is uneven and $b_3(p)$ is even in momentum, from
various mean values with the momentum-even distributions only the following
terms remain nonzero
\be
{\omega_c\over 2 m_e}\sum\limits_p \begin{pmatrix}f\cr g \end{pmatrix}  \V q\cdot (\V e_z\times \V p )\V b_\perp(p)&=&{\omega_c
  \over D}\V b_\perp (q)\times \V e_z\begin{pmatrix}E_f\cr E_g\end{pmatrix}
\nonumber\\
\sum\limits_p \begin{pmatrix}f\cr g \end{pmatrix}{(\V p\cdot \V q)^2\over 2 m_e^2}&=&{q^2\over D m_e}\begin{pmatrix}E_f\cr E_g\end{pmatrix}
\ee
where $D$ denotes the dimension and the mean (polarization) kinetic energy is denoted as
\be
\begin{pmatrix}E_f\cr E_g\end{pmatrix}=\sum\limits_p \begin{pmatrix}f\cr g \end{pmatrix} {p^2\over 2m_e}.
\ee 
Here and in the following we use  $\V q\V {\partial_p}
b(p)\approx b(q)$ strictly valid only for linear spin-orbit coupling and
neglect higher-order moments than $o(p^2 b(p))$.
Besides (\ref{Bg}) we will use further shorthand notations
\ba
B_{g3}=\sum_p b_3 g,\quad B_{gii}=\sum_p b_{\perp i}^2 g,\quad \V B_\perp=\V b_{\perp}(q)
\label{Bg3}
\end{align}
and analogously for $g\leftrightarrow f$.

The equation for the density dipole-excitation is given by the first line of (\ref{final}) and one needs the expansion of the polarizations
\be
\Pi_0={n q^2\over m_e \omega_+^2}+o(q^3)
\ee
and analogously
\be
\V \Pi={\V m q^2\over m_e \omega_+^2}+o(q^3).
\label{VP}
\ee
 This means for neutral scattering the combinations $V_0 \Pi_0$ and $V_0 \V \Pi$ vanish.

According to (\ref{abbrev}) we need the expansion of (\ref{Pg})
\ba
&\V \Pi_g=-i\left \{\V e_z \left [s_0-{B_g^2\over 2}
    (1-\Sigma_n\partial_{\Sigma_n})+B_{g3}\Sigma_n\partial_{\Sigma_n}
\right .\right .\nonumber\\&\left .\left .+{B_{g33}\over
      2} \Sigma_n^2\partial_{\Sigma_n}^2+{q^2E_g\over D m}\partial_\omega^2 
\right ]
-\V B_\perp\times \V e_z {\omega_c E_g\over D}\partial_\omega^2
\right \} {2\omega\over \omega^2-4\Sigma_n^2}
\end{align}
and also $+o(q^3)$
\ba
&\V \Pi_{xf}=i\left \{\V e_z \left [
{q^2\over m} (n-{B_f^2\over
  2}(1-\Sigma_n\partial_{\Sigma_n})+B_{f3}\Sigma_n\partial_{\Sigma_n}
\right . \right .\nonumber\\&\left . \left . 
+{B_{f33}\over
  2}\Sigma_n^2\partial_{\Sigma_n}^2)\partial_\omega -n
B_{\perp 3}\Sigma_n\partial_{\Sigma_n}-B_{\perp 3}B_{f3}
\Sigma_n^2\partial_{\Sigma_n}^2\right ]
\right .\nonumber\\&\left .
-\V B_\perp (n-B_{f3}
   (1-\Sigma_n\partial_{\Sigma_n}))
\right .\nonumber\\&\left .
-\V B_\perp\times \V
 e_z B_{\perp 3} {\omega_c E_g\over D}(1-\Sigma_n\partial_{\Sigma_n}\partial_\omega
\right \} {2\Sigma_n\over 4 \Sigma_n^2-\omega^2}
\end{align}
such that we have the precession term $\V \Pi_2=\V \Pi_{xf}+\V \Pi_{g}$.

For $\Pi_3$ we need besides (\ref{VP}) according to (\ref{abbrev})
\ba
&\V \Pi_{xg}=i\left \{
\V e_z\times \V B_\perp (s_0+B_{g3} \Sigma_n\partial_{\Sigma_n})
\right .\nonumber\\&\left .
+\left [\V B_\perp B_{\perp 3}-\V e_z B_\perp^2\right ] {\omega_c E_g\over D} \partial_\omega^2 \right \} {2\Sigma \over 4\Sigma^2-\omega^2}
\end{align}
and 
\ba
&\V \Pi_{e}=\left \{
\V B_\perp
(s_0-B_{g3}(1-\Sigma_n\partial_{\Sigma_n})-B_{g33}\Sigma_n\partial_{\Sigma_n})
\right .\nonumber\\&\left .
+\V
b_\perp(q)\times \V e_z {\omega_c E_g\over D} \partial_omega^2
\right\} {4\Sigma_n^2\over \omega(4\Sigma_n^2-\omega^2) }
\end{align}
such that
$
\V \Pi_3=o(q).
$
The in-plane terms are a little bit more lengthy 
\ba
&\overleftrightarrow \Pi_{xe}=
2 \left \{
\begin{pmatrix} 1&0&0\cr 0&1&0\cr 0&0&0\end{pmatrix}
\left (
s_0+{q^2 E_g\over m D}\partial_\omega^2+(B_{g3}+{B_g^2\over 2})
\Sigma_n\partial_{\Sigma_n}
\right .\right .\nonumber\\&\left .\left .+{B_{g33}\over
    2}\Sigma_n^2\partial_{\Sigma_n}^2\right )
+
\begin{pmatrix} 0&0&B_{\perp 2}\cr 0&0&-B_{\perp 1}\cr B_{\perp 2}&-B_{\perp 1}&0\end{pmatrix}
{\omega_c E_g\over D}\partial_\omega^2
\right .\nonumber\\&\left .
-
\begin{pmatrix} B_{g11}&0&0\cr 0&B_{g22}&0\cr 0&0&-B_g^2\end{pmatrix}
\right \}
{2\Sigma_n\over 4 \Sigma_n^2-\omega_+^2}
\end{align}
and
\ba
&\overleftrightarrow \Pi_{fe}=
\left \{
-B_{\perp 3} \begin{pmatrix} 1&0&0\cr 0&1&0\cr 0&0&0\end{pmatrix}
\left (
n\Sigma_n\partial_{\Sigma_n}+B_{f3}\Sigma_n^2\partial_{\Sigma_n}^2\right ) 
\right .\nonumber\\&\left .
+\begin{pmatrix} 2 B_{\perp 1} B_{\perp 2}&B_{\perp 2}^2-B_{\perp 1}^2&B_{\perp 2}\cr
  B_{\perp 2}^2-B_{\perp 1}^2&-2 B_{\perp 1} B_{\perp 2}&-B_{\perp 1}\cr
  B_{\perp 2}&-B_{\perp 1}&0\end{pmatrix}{\omega_c E_f\over D}\partial_\omega^2
\right .\nonumber\\&\left .
+ \left [\begin{pmatrix}
 1&0&0\cr 0&  1&0\cr 0&0&0\end{pmatrix}\left [\left
   (B_{f3}+{B_f^2\over2}\right )\Sigma_n\partial_{\Sigma_n}
+{B_{f33}\over 2}
\Sigma_n^2\partial_{\Sigma_n}^2\right ]
\right .\right .\nonumber\\&\left .\left .
+\begin{pmatrix}
  n-B_{f11}&0&0\cr 0&  n-B_{f22}&0\cr 0&0&B_f^2\end{pmatrix}\right ]{q^2\over m}\partial_\omega
\right \}
{4\Sigma_n^2\over \omega(4 \Sigma_n^2-\omega_+^2)}
\end{align}

The terms coming from the Mermin relaxation time become
\ba
\Pi_{0\mu}&=-{\p \mu n\over \omega_+}-{n q^2\over m_e
  \omega_+^3}+o(q^3,b^3,\sigma_nb(q)^2)
\nonumber\\
\V \Pi_\mu&=-{\p \mu \V s \over \omega_+}-{s q^2\over m_e
  \omega_+^3}\V e_z+o(q^3,b^3,\sigma_nb(q)^2)
\label{mermin}
\end{align}
where we use (\ref{s}). The terms $\V \Pi_{\mu e}$ and $\V \Pi_{x\mu}$ vanish
at this level of expansion.

\subsection{Long wave length expansion in quasi 2D systems}

In the cases discussed in this paper we are not interested in 3D spin-orbit coupling
such that we can neglect the terms $b_3$. 
Summarizing the results of the last chapter we obtain for
the coupled dispersion (\ref{final}) the terms
\be
\Pi_0={n q^2\over m_e \omega_+^2},\quad \V \Pi={\V s q^2\over m_e
  \omega_+^2},\quad s=s_0-{B_g^2\over 2},
\ee
\ba
&\V \Pi_2=i\V e_z \left \{
\left [-s_0+{B_g^2\over 2}(1-\Sigma_n\partial_{\Sigma_n})-
{q^2E_g\over D m_e}\partial_\omega^2\right ]{2\omega\over \omega^2-4\Sigma_n^2} 
\right .\nonumber\\ &\left .
-\V B_\perp\times \V e_z {\omega_c E_g\over D}\partial_\omega^2
+{q^2\over m} (n-{B_f^2\over
  2}(1-\Sigma_n\partial_{\Sigma_n})) {2\Sigma_n\over 4 \Sigma_n^2-\omega^2}
\right \}
\nonumber\\ &
+i\V B\times \V
 e_z {\omega_c E_g\over D}\partial_\omega^2
 {2\omega\over \omega^2-4 \Sigma_n^2}
-\V B_\perp n {2\Sigma_n\over 4 \Sigma_n^2-\omega^2},
\end{align}
\ba
&\V \Pi_3={\V s q^2\over m_e
  \omega_+^2}
-i\left [
\V e_z\times \V B_\perp s_0-\V e_z B_\perp^2 {\omega_c E_g\over D} \partial_\omega^2 \right ] {2\Sigma \over 4\Sigma^2-\omega^2}
\nonumber\\ &
+\V B_\perp s_0 {4\Sigma_n^2\over \omega(4\Sigma_n^2-\omega^2) },
\end{align}
and
\ba
&\overleftrightarrow \Pi=
\left \{
\begin{pmatrix} 1&0&0\cr 0&1&0\cr 0&0&0\end{pmatrix}
\left (
s_0+{q^2 E_g\over m D}\partial_\omega^2+{B_g^2\over 2}
\Sigma_n\partial_{\Sigma_n}\right )
\right . \nonumber\\ &\left .
+\begin{pmatrix} -B_{g11}&0&B_{\perp 2}{\omega_c E_g\over D}\partial_\omega^2\cr 0&-B_{g22}&-B_{\perp 1}{\omega_c E_g\over D}\partial_\omega^2\cr B_{\perp 2}{\omega_c E_g\over D}\partial_\omega^2&-B_{\perp 1}{\omega_c E_g\over D}\partial_\omega^2&-B_g^2\end{pmatrix}
\right \}
\nonumber\\ &
\times {4\Sigma_n\over 4 \Sigma_n^2-\omega_+^2}
\nonumber\\ &
+\left \{
\begin{pmatrix} 2 B_{\perp 1} B_{\perp 2}&B_{\perp 2}^2-B_{\perp 1}^2&B_{\perp 2}\cr
  B_{\perp 2}^2-B_{\perp 1}^2&-2 B_{\perp 1} B_{\perp 2}&-B_{\perp 1}\cr
  B_{\perp 2}&-B_{\perp 1}&0\end{pmatrix}{\omega_c E_f\over D}\partial_\omega^2
\right . \nonumber\\ &\left .
+\begin{pmatrix}
  n+{B_f^2\over2}\Sigma_n\partial_{\Sigma_n}-B_{f11}&0&0\cr 0&  n+{B_f^2\over2}\Sigma_n\partial_{\Sigma_n}-B_{f22}&0\cr 0&0&B_f^2\end{pmatrix}
\right . \nonumber\\ &\left .
\times {q^2\over m_e}\partial_\omega
\right \}
{4\Sigma_n^2\over \omega(4 \Sigma_n^2-\omega_+^2)}
\end{align}
together with (\ref{mermin}).

\bibliography{bose,kmsr,kmsr1,kmsr2,kmsr3,kmsr4,kmsr5,kmsr6,kmsr7,delay2,spin,spin1,refer,delay3,gdr,chaos,sem3,sem1,sem2,short,cauchy,genn,paradox,deform,shuttling,blase,spinhall,spincurrent,tdgl,pattern,zitter}

\end{document}